\documentclass{jfm}
\usepackage{graphicx}
\usepackage[section]{placeins}
\usepackage{float}
\usepackage{epstopdf, epsfig}
\usepackage[utf8]{inputenc}
\usepackage{textcomp}
\usepackage{multirow}
\usepackage{graphicx}
\usepackage{epstopdf}
\usepackage{pdfpages}
\usepackage{soul}
\usepackage{xcolor}
\usepackage{natbib}
\usepackage{subfigure}
\usepackage{hyperref}
\hypersetup{
	colorlinks = true,
	linkcolor  = blue,
	urlcolor   = blue,
	citecolor  = blue,
}

\shorttitle{Drag modification by finite-size particles}
\shortauthor{C. Wang, L. Yi, L. Jiang and C. Sun}

\title{How do the finite-size particles modify the drag in Taylor-Couette turbulent flow}

\author{Cheng Wang\aff{1}, Lei Yi\aff{1}, Linfeng Jiang\aff{1} \and Chao Sun\aff{1,}\aff{2}\corresp{\email{chaosun@tsinghua.edu.cn}}}

\affiliation{
	\aff{1}Center for Combustion Energy, Key Laboratory for Thermal Science and Power Engineering of Ministry of Education, Department of Energy and Power Engineering, Tsinghua University, Beijing 100084, China
	\aff{2}Department of Engineering Mechanics, School of Aerospace Engineering, Tsinghua University, Beijing 100084, China}

\begin{document}
	\maketitle

\begin{abstract}
\textcolor{black} {	We experimentally investigate the drag modification by neutrally buoyant finite-size particles with various aspect ratios in a Taylor-Couette (TC) turbulent flow. The current Reynolds number, $Re$, ranges from $6.5\times10^3$ to $2.6\times10^4$, and the particle volume fraction, $\Phi$, is up to $10\%$. Particles with three kinds of aspect ratio, $\lambda$, are used: $\lambda=1/3$ (oblate), $\lambda=1$ (spherical) and $\lambda=3$ (prolate). Unlike the case of bubbly TC flow \citep{van2013importance, verschoof2016bubble}, we find that the suspended finite-size particles increase the drag of the TC system regardless of their aspect ratios. The overall drag of the system increases with increasing $Re$, which is consistent with the literature. In addition, the normalized friction coefficient, $c_{f,\Phi}/c_{f,\Phi=0}$, decreases with increasing $Re$, the reason could be that in the current low volume fractions the turbulent stress becomes dominant at higher $Re$. The particle distributions along the radial direction of the system are obtained by performing optical measurements}
\textcolor{black} { \textcolor{black}at $\Phi=0.5\%$ and $\Phi = 2\%$.}
\textcolor{black} {As $Re$ increases, the particles distribute more evenly in the entire system, which results from both the greater turbulence intensity and the more pronounced finite-size effects of the particles at higher $Re$. Moreover, it is found that the variation of the particle aspect ratios leads to different particle collective effects. The suspended spherical particles, which tend to cluster near the walls and form a particle layer, significantly affect the boundary layer and result in maximum drag modification. The minimal drag modification is found in the oblate case, where the particles preferentially cluster in the bulk region, and thus the particle layer is absent.}
 {Based on the optical measurement results, it can be concluded that, in the low volume fraction ranges ($\Phi=0.5\%$ and $\Phi = 2\%$ here),}
\textcolor{black} { the larger drag modification is connected to the near-wall particle clustering. The present findings suggest that the particle shape plays a significant role in drag modification, and the collective behaviors of rigid particles provide clues to understand the bubbly drag reduction.}
\end{abstract}

\begin{keywords} Taylor-Couette, turbulence, multiphase flow
\end{keywords}

\section{Introduction}\label{sec:intro}
\textcolor{black} {Particle-laden flow is ubiquitous both in nature and industry, including plankton in the ocean \citep{pedley1992hydrodynamic}, dust and virus dispersed in the atmosphere \citep{mittal2020flow} and catalytic particles in industrial settings \citep{wang2019self, wang2020experimental}. For the suspension of rigid spheroidal particles, the system can be characterized by the Reynolds number of the flow, $Re$, and the particle parameters, which include the density, $\rho_p$, the diameter, $d_p$,  the aspect ratio, $\lambda$, and the volume fraction of the particle, $\Phi$. According to the ranges of $d_p$ and $\Phi$, the mechanism of the fluid-particle interaction can be roughly classified into two types: (\romannumeral1) one-way coupled and (\romannumeral2) two-/four-way coupled \citep{elghobashi1994predicting, voth2017anisotropic}. When $d_p \ll \eta_K$ (the dissipative length scale of the flow) and $\Phi$ is low, the fluid and the particles can be treated as one-way coupled, hence the fluid is unaffected. In this regime, the particle-laden flow can be described by the point-particle model, which has been extensively verified and used in the simulation studies \citep{saw2008inertial, rusconi2014bacterial, frankel2016settling, park2018rayleigh, lohse2018bubble, lovecchio2019chain}.}
 {For $d_p \ll \eta_K$ but $\Phi$ is high enough, the particles can still be modeled as point-particles but would instead affect the surrounding fluid since the fluid and particles are two-/four-way coupled now.}
\textcolor{black} {However, when $d_p$ exceeds $\eta_K$ (typically $d_p > 10\eta_K$ \citep{voth2017anisotropic}), the particles, which are referred to as finite-size particles, can modify the surrounding flow field by two-way or four-way coupling way. To resolve the flow field around the finite-size particles, experiments and fully-resolved simulations, which have been conducted to study the physics of the finite-size effects and the particle dynamics \citep{magnaudet2000motion, peskin2002immersed, uhlmann2008interface, tagawa2013clustering, wang2017fully, bakhuis2018finite, jiang2020rotation,will2021kinematics, assen2021strong}, are required.}

\textcolor{black} {The finite-size effects on the microscopic scale can cause turbulence modulation on the macroscopic scale. It has been reported that tremendous drag reduction can be achieved by a small amount of large bubble injection \citep{van2005drag, van2007bubbly, van2013importance, verschoof2016bubble, ezeta2019drag}, however, of which the mechanism is still not fully understood. The main difficulty in studying the bubbly flow in the experiments is the deformability of the bubble, which makes it difficult to fix the bubble shape and size. On the contrary, through performing density matching, suspended rigid spheroidal particles can be used to partially solve the difficulty above, making it possible to delve into the mechanism of bubbly drag reduction.}

\textcolor{black} {Indeed, recent studies have hinted that the spherical particles are promising in achieving drag modification both in channel flow \citep{lashgari2014laminar, picano2015turbulent} and duct flow \citep{zade2018experimental}.}
\textcolor{black} {However, on the one side, few studies have directly measured the drag of the flow, under what conditions the particles will increase or decrease the turbulence drag is still unknown. \cite{bakhuis2018finite} experimentally investigated the effects of finite-size rigid spherical particles on the drag of a Taylor-Couette turbulent flow, where they found that the overall drag does not vary much in their explored high Reynolds number range and the low particle volume fraction range. While in a later study, \cite{ardekani2019turbulence} varied the aspect ratios of the suspended particles in wall-bounded turbulence and found turbulence attenuation (compared to the single-phase flow) in prolate and oblate cases. In their study, an overall drag reduction was also found for the oblate particle with $\lambda\leq1/3$ when compared to the single-phase case.}
\textcolor{black} {It should be noted that turbulence attenuation is not necessary to connect to an overall drag reduction and vice versa. For example, the turbulence attenuation but overall drag enhancement might be achieved at the same time at the cost of high particle volume fractions since the particle-induced stress is greatly enhanced \citep{picano2015turbulent}.}
\textcolor{black} {Given the importance of the boundary layer for turbulent stress and drag, an explanation was proposed by \cite{ardekani2019turbulence} for the mechanism of turbulence modulation in terms of the near-wall dynamics of different particles. On the other side, studies in recent years have shed some light on the physics of the turbulence modulation induced by particles. For example, \cite{ardekani2017drag} performed simulations in turbulent channel flow laden with oblate particles up to a volume fraction of $\Phi=15\%$, and they have observed an overall drag reduction. In their study, two possible mechanisms are found to be responsible for the particle-induced drag reduction: (\romannumeral1) the absence of the near-wall particle layer and (\romannumeral2) the alignment of the major axes of the oblate particles to the wall. Besides, experimental findings have also suggested that the collective effects of the particles and bubbles can significantly affect the flow properties \citep{calzavarini2008quantifying, colin2012turbulent, van2013importance, maryami2014bubbly, almeras2017experimental, mathai2018dispersion}.}

\textcolor{black} {Among various flow geometries, a closed setup, where drag can be directly measured, is convenient for evaluating the drag modification of the flow induced by suspended particles. In this work, we employ a TC system \citep{grossmann2016high} - the flow between two coaxial cylinders - to suspend the particles. The Reynolds number, $Re$, the particle volume fraction, $\Phi$, and the particle aspect ratio, $\lambda$, are varied to study their effects on the drag. To find out the mechanism of the drag modification, optical measurements are performed to obtain the particle distribution in the radial direction of the system.}

\begin{figure}
	\centering
	\includegraphics[width=1\linewidth]{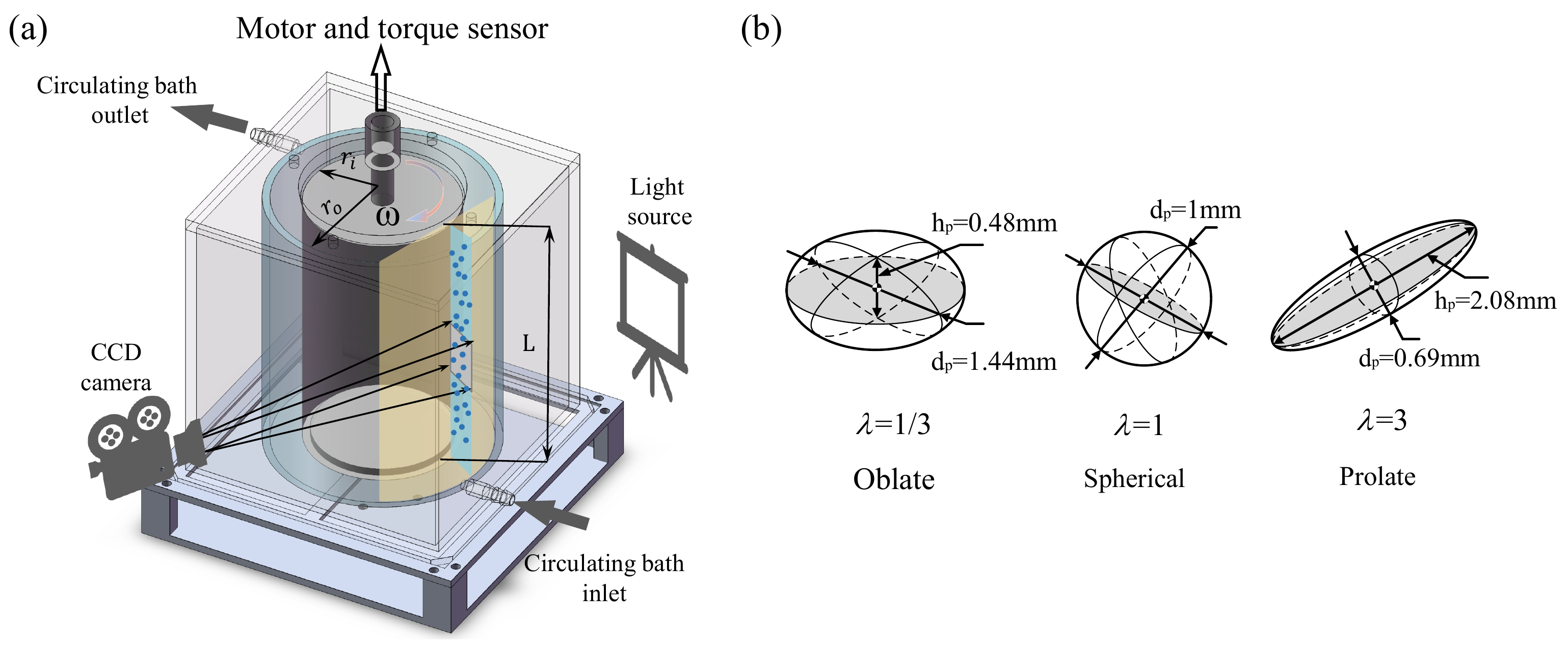}
	\vspace{-5 mm}
	\caption{\label{fig1_sketch} $(a)$ The sketch of the Taylor-Couette facility. The gap between the two coaxial cylinders is filled with glycerine-water solution to match the particle density. The neutrally buoyant particles (blue dots) are recorded by the CCD camera. The outer cylinder is surrounded by a PMMA cubic circulating water bath, which keeps the temperature of the system at $T=22\pm0.1$\ \textcelsius\ during the experiments. $(b)$ shows the sketches and sizes of the particles used in this work.}
\end{figure}
\section{Experiment}\label{sec:setup}
\subsection{Experimental facility}

\textcolor{black} {The TC facility used in the present work is shown by the sketch in figure~\ref{fig1_sketch}(a). The inner cylinder is made of aluminum, while the outer one is made of PMMA (polymethyl methacrylate) to perform optical measurements. The outer cylinder is surrounded by a PMMA cubic circulating bath, which keeps the system temperature at $T= 22\pm 0.1$\ \textcelsius\ during the experiments. The radius of the inner and outer cylinder is $r_i$ = 25\ mm and $r_o$ = 35\ mm, respectively, giving the gap width $d$ = $r_o-r_i$ = 10\ mm and the radius ratio $\eta=r_i/r_o=0.714$. The height of the inner cylinder is $L$ = 75\ mm, which gives the aspect ratio of the system $\Gamma=L/d=7.5$.}

\textcolor{black} {During the experiments, the outer cylinder is fixed, while the inner cylinder is connected to and driven by a rheometer (Discovery Hybrid Rheometer, TA Instruments), and the overall torque of the system is measured at the same time by the torque sensor of the rheometer. The control parameter of TC flow is the Reynolds number defined by:
\begin{equation}\label{Re}
	Re=\frac{\omega_ir_id}{\nu},
\end{equation}
where $\omega_i$ is the angular velocity of the inner cylinder and $\nu$ is the kinematic viscosity of the fluid. In the present work, the $\omega_i$ ranges from $50$ to $200$\ rad/s, which gives the $Re$ ranges from $6.5\times10^3$ to $2.6\times10^4$.}

\textcolor{black} {The overall torque of the system, $\tau_{total}$, that needed to drive the inner cylinder at a constant angular velocity $\omega_i$, can be divided into two parts: (\romannumeral1) the torque, $\tau$, due to the sidewall of the inner cylinder (TC flow), (\romannumeral2) the torque, $\tau_{end}$, contributed from the von K\'{a}rm\'{a}n flow at the end plates. Since the aspect ratio of the system is relatively small here, the rotation of the end plates generates secondary vortices that contribute to the torque measurements \citep{bagnold1954experiments, hunt2002revisiting}. The calibration of the end effect due to the von k\'{a}rm\'{a}n flow between the bottom and top lids of the inner and outer cylinder can be found in Appendix~\ref{sec:appendixA}. Here, only $\tau$ is used, and it can be non-dimensionalized as:
\begin{equation}\label{G}
	G=\frac{\tau}{2\pi L\rho_f\nu^2},
\end{equation}
where $L$ is the length of the inner cylinder and $\rho_f$ is the fluid density. }

\textcolor{black} {In the experiments, the gap between the bottom and top lids of the inner and outer cylinders is 2\ mm and larger than the particle diameter, the particles therefore would enter in and escape from the bottom and top gaps between the inner and outer cylinders.}
\textcolor{black} {For each experiment, the system is initially started from a static state, and then the inner cylinder is imposed a constant angular velocity by the motor of the rheometer. Before measuring the torque, the inner cylinder is kept rotating to ensure the establishment of a statistically stationary state. For each $\omega_i$, the torque measurements are repeated 3 times, and the averaged value is used, of which the standard deviations are less than $1\%$.}

\subsection{Preparation of the finite-size particles}
\textcolor{black} {The particles we used are printed with a 3D printer using photosensitive resin (Elastic Resin of Formlabs. Inc) and are then cured at 60\ \textcelsius \ for 60 mins. To check their water-absorbing ability, the particles are immersed in the static glycerin-water solution for 2 hours, and afterward the change in weight is found to be negligible.}
\textcolor{black} {The density of the particle is $\rho_p=1.06 \times 10^3$\ kg/m$^{3}$ (averaged value based on more than 1000 particles). The particle volume fraction, $\Phi$, ranges from $0\%$ to $10\%$ with a spacing of $2\%$.}
\textcolor{black} {In addition, optical measurements and additional torque measurements are performed at $\Phi=0.5\%$.}
\textcolor{black} {To study the shape effect of the finite-size particles, we change the particle aspect ratio, $\lambda=h_p/d_p$, where $d_p$ is the length of the symmetric axis of the particle, and $h_p$ is the length that perpendicular to it. In this work, $\lambda$ is designed to be equal to 1/3, 1, and 3, corresponding to the oblate, spherical and prolate, respectively (see figure~\ref{fig1_sketch}(b)). We fix the volume of each particle the same for three cases of aspect ratio. The equivalent diameter of the particle based on the volume, $d_e$, is equal to the diameter of the spherical particle, i.e., $d_e=d_s=1$\ mm. Note that the particles are verified to be finite-size particles. For example, the dissipative length scale of the TC flow, $\eta_K$, can be estimated as around $0.075$\ mm when $\omega_i=50$\ rad/s, which gives}
\textcolor{black} {$d_e\textgreater10\eta_K$}
\textcolor{black} {and satisfies the assumption of the finite-size model \citep{voth2017anisotropic}. }

\textcolor{black} {To eliminate the effect of the gravity and the buoyancy force, the density of the particle and the fluid are matched using glycerin-water solution (the weight fraction of the glycerin is $w_t=25\%$). The solution density is $\rho_f=1.0598\times 10^3$\ kg/m$^{3}$ and the kinematic viscosity is $\nu=1.913\times 10^{-6}$\ m$^{2}$/s at $T=22$\ \textcelsius \  \citep{cristancho2011volumetric}, giving a 0.02\% density mismatch from the particle density. Since the temperature of the system is well controlled by the circulating water bath, the effects of the temperature variation on the density and viscosity of the solution can be neglected in the present work.}

\textcolor{black} {The particle motion is recorded by a CCD camera (MD028MU-SY, Ximea.Inc) at a frequency $f_c=15$ Hz, and then the particle positions are obtained by performing particle detection. The flow domain is illuminated by a light source, which is shown in figure~\ref{fig1_sketch}(a). The details of the detection methods and examples can be found in Appendix~\ref{sec:appendixC}.}

\begin{figure}
	\centering
	\includegraphics[width=1\linewidth]{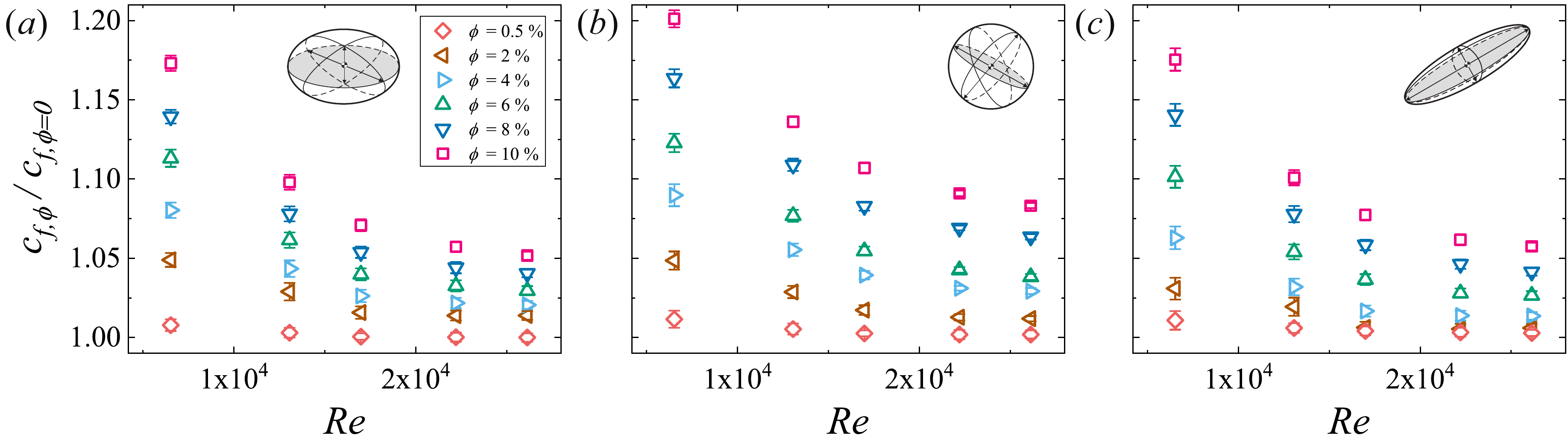}
	\vspace{-5 mm}
	\caption{\label{fig2_cf} The normalized friction coefficient of the Taylor-Couette system, $c_{f,\Phi}/c_{f,\Phi=0}$, for (a) $\lambda=1/3$ (oblate), (b) $\lambda=1$ (spherical) and (c) $\lambda=3$ (prolate), where $c_{f,\Phi=0}$ is the friction coefficient for the single-phase case. The accuracy of the experiments is indicated by the error bars, which are typically less than 1\%.}
\end{figure}
\section{Results}\label{sec:result}
\subsection{Drag modification}

\textcolor{black} {Firstly, we study the effects of changing the Reynolds number on the drag of the system. For the global analysis of the drag modification by bubbles or particles, the friction coefficient, $c_{f,\Phi}$, which evaluates the skin frictional drag of the flow system, has been extensively used in various flow geometries \citep{sanders2006bubble, verschoof2016bubble, olivucci2021reduction}. In the TC system, the friction coefficient is defined as \citep{lathrop1992transition, van2011torque}:
\begin{equation}\label{cf}
	c_{f,\Phi}= \frac{(1-\eta)^2}{\pi \eta^2}\frac{G}{Re^2}.
\end{equation}
Figure~\ref{fig2_cf} shows the normalized friction coefficients of the system, $c_{f,\Phi}/c_{f,\Phi=0}$, where $c_{f,\Phi=0}$ is the friction coefficient of the single-phase case. It is found that the drag modification increases with increasing $\Phi$ for a given $Re$. In the investigated ranges, the greatest drag enhancement is found to be around $20\%$, which is observed in the spherical case at $Re=6.5\times10^3$ and $\Phi=10\%$ (figure~\ref{fig2_cf}(b)). Additionally, $c_{f,\Phi}/c_{f,\Phi=0}$ decreases as the $Re$ increasing, which has also been found and is referred to as the shear-thinning effect in the suspension of deformable particles \citep{adams2004influence} and the emulsions \citep{yi2021global, rosti2021shear}.}
\textcolor{black} {Note that, the overall drag of the system ($G$ or $\tau$, which are not shown here), increases with increasing $Re$ and is consistent with the literature \citep{stickel2005fluid, fall2010shear, picano2013shear}.}

\textcolor{black} {The modification of the friction coefficient is related to the changes of the wall stress, which can be decomposed into three parts when no external force and torque is applied on particles:}
\textcolor{black} {
\begin{equation}\label{tauw}
	\tau_{w,\Phi} = \tau_{v}+\tau_{T}+ \tau_{p}= \tau_{v}+\tau_{T_f}+\tau_{T_p} +\tau_{p},
\end{equation}
where $\tau_{v}$ the viscous stress of the fluid phase, $\tau_{p}$ the particle-induced stress, and $\tau_{T}= \tau_{T_f}+\tau_{T_p}$ the turbulent Reynolds stress of the combined phase, $\tau_{T_f}$ and $\tau_{T_p}$ the turbulent Reynolds stress of fluid and particulate phase, respectively. Following the studies in \cite{zhang2010physics} and \cite{picano2015turbulent}, the explicit expression of each term above in a Couette flow \citep{batchelor1970stress, wang2017modulation} can be written as:
\begin{equation}\label{taufv}
	\tau_{v} = \mu (1-\Phi) \frac{\mathrm{d}U_f}{\mathrm{d}r},
\end{equation}
\begin{equation}\label{tauft}
	\tau_{T_f} = -\rho(1-\Phi) \langle u^{\prime}_fv^{\prime}_f \rangle,
\end{equation}
\begin{equation}\label{taupt}
	\tau_{T_p} = -\rho \Phi \langle u^{\prime}_pv^{\prime}_p \rangle,
\end{equation}
\begin{equation}\label{taups}
	\tau_{p} = \Phi \langle \sigma^p_{xy} \rangle,
\end{equation}
where $\mu$ the dynamic viscosity of the fluid, $U_f$ the mean fluid velocity in the azimuthal direction, $u^{\prime}$ and $v^{\prime}$ the velocity fluctuation in the azimuthal and radial direction with the subscripts $f$ and $p$ denoting fluid and particulate phase, respectively, $\sigma^p_{xy}$ the general stress in the particulate phase, normal to the cylindrical surface and pointing in the radial direction. Here we note that the contribution to fluid viscous stress due to velocity in the axial direction is neglected since it is a minor effect.}

\textcolor{black} {At a fixed $Re$, the wall stress of single-phase case, $\tau_{w,\Phi=0} = \mu\frac{\mathrm{d}U}{\mathrm{d}r}|_{r=r_i}$, is a constant, and the particle-induced stress ($\tau_{p}$) increases with increasing $\Phi$. In addition, it has been reported that in channel flow \citep{picano2015turbulent}, for a fixed $Re$ in the current low particle volume fraction ranges, the viscous stress ($\tau_{v}$) weakly depends on $\Phi$, and the turbulent Reynolds stress of the combined phase ($\tau_{T}$) increases with increasing $\Phi$. Though there are differences between the channel flow and the TC flow, the result in the channel flow is instructive for the understanding of the current study. Considering the dependence above of the stress on $\Phi$ at a fixed $Re$, the normalized friction coefficient increases with increasing $\Phi$, which is consistent with the trends in figure~\ref{fig2_cf}. On the other hand, for a fixed low $\Phi$ (as in this work), the turbulent Reynolds stress of fluid phase in single-phase flow ($\tau_{T_f,\Phi=0}$) increases greatly as $Re$ increases, which makes the contribution of particulate phase ($\tau_{T_p}$ and $\tau_{p}$) relatively insignificant and therefore results in the decreasing trends of normalized friction coefficients with increasing $Re$. Another feature of the normalized friction coefficients in figure \ref{fig2_cf} is that, as $Re$ increases, the differences between various $\Phi$ decrease. Specifically, when $Re$ is small, the contributions of particulate phase to the wall stress are non-negligible, hence the normalized wall stress (friction coefficients) highly depends on $\Phi$ through $\tau_{T_p}$ and $\tau_{p}$. However, when $Re$ is high enough and $\Phi$ is low (as in this work), the turbulent Reynolds stress due to fluid phase in particle-laden flow and single-phase flow ($\tau_{T_f}$ and $\tau_{T_f,\Phi=0}$) plays the dominant role while $\tau_{T_p}$ and $\tau_{p}$ become minor, therefore the contributions from particles become less and less important as $Re$ increases. Hence, the effect of changing $\Phi$ on the normalized friction coefficients becomes smaller at higher $Re$ in the current parameter regimes.}

\textcolor{black} {On the other side, taking the particle-laden flow as a continuous effective fluid, the interaction between the particles and the resulting particle distributions are related to the rheology of a particle-laden flow, which can be quantified by the effective viscosity, $\nu_{eff}$. In the Stokes regime ($Re\ll1$), the rheology of dense granular suspensions has been extensively investigated, and the dependence of the effective viscosity on the particle volume fraction has been discussed \citep{guazzelli2018rheology}.}
\textcolor{black} {In the current semi-dilute regime ($\Phi\leq10\%$), the normalized effective viscosity of the granular flow in the Stokes regime, $\nu^S_{eff}$, can be approximated using the Krieger \& Dougherty (K-D) formula \citep{krieger1959mechanism}:
\begin{equation}\label{KD}
	{\nu^S_{eff}\over{\nu_f}} =(1-\frac{\Phi}{\Phi_m})^{-[\eta]\Phi_m},
\end{equation}
where $\nu_f$ the viscosity of the fluid, $\Phi_m$ the maximum packing particle fraction ($\Phi_m = 0.65$ is used here as done in \cite{stickel2005fluid}) and $[\eta]={5\over2}$ for rigid spheres.}
\textcolor{black} {In this work, the $\nu_{eff}$ was calculated using the same method as in previous works \citep{bakhuis2021catastrophic, yi2021global}, and the details can be found in Appendix~\ref{sec:appendixB}. As shown in figure~\ref{fig3_effectivevis}, we compare the experimental results with the K-D formula. Though the model captures the increasing trend of the effective viscosity, the $Re$-dependence of the effective viscosity is missing, indicating that the relation in the Stokes regime is no longer valid in the current situation. The disparity found between the experimental data and the model can be understood since equation~\ref{KD} was obtained based on the assumptions that}
\textcolor{black} {the flow is inertialess}
\textcolor{black} {and the particles distribute uniformly in the Stokes regime. However, for the current turbulent flow ($Re\sim10^4$) and the suspended finite-size particles, the particle inertia is non-negligible and can be measured by the Stokes number, which is far beyond unity as shown in the later section. Moreover, as has been reported in previous numerical studies \citep{picano2015turbulent, ardekani2019turbulence}, the particles show collective effects, which are also found in our experiments and will be discussed later. Note that when $Re=6.5\times10^3$, the flow could be in the classical regime with laminar boundary layers and a turbulent bulk with Taylor vortex \citep{grossmann2016high}, which accounts for the more rapidly increasing trend than the model and other experimental data. Using the effective viscosity, the shear-thinning effect of the particle-laden flow is also examined, which can be well described by the Herschel–Bulkley model \citep{herschel1926konsistenzmessungen}, and the details can be found in Appendix~\ref{sec:appendixB}.}
\begin{figure}
	\centering
	\includegraphics[width=1\linewidth]{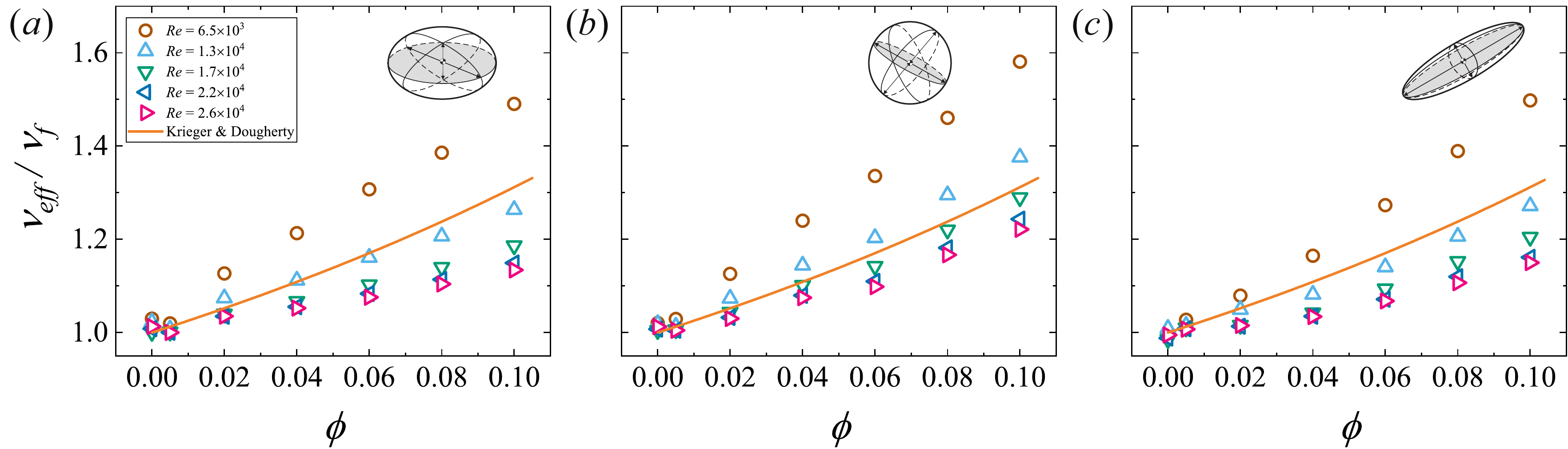}
	\vspace{-5 mm}
	\caption{\label{fig3_effectivevis} The non-dimensional effective viscosity, $\nu_{eff}/\nu_f$, as a function of the particle volume fraction, $\Phi$, for (a) $\lambda=1/3$ (oblate), (b) $\lambda=1$ (spherical) and (c) $\lambda=3$ (prolate), where $\nu_f$ is the kinematic viscosity of the solution.}
\end{figure} 

\subsection {Effects of the particle aspect ratio}

\textcolor{black} {Next, we investigate the effects of changing the particle aspect ratio on the drag of the system. Particles with three kinds of aspect ratio are used: $\lambda=1/3$ (oblate), $\lambda=1$ (spherical) and $\lambda=3$ (prolate). The results are already given in figure~\ref{fig2_cf} and figure~\ref{fig3_effectivevis}. It is found that the particles could increase the drag of the system regardless of their aspect ratio, which is attributed mostly to the hydrodynamics interactions and rarely to the frictional contact between the particles in the current low $\Phi$ range \citep{guazzelli2018rheology}. The shear-thinning effect and the increasing trend of the drag modification with increasing $\Phi$ are found to be ubiquitous in all cases.}

\textcolor{black} {It is remarkable, however, that particles with different aspect ratios could affect the drag of the system to different degrees, even for the same $Re$ or $\Phi$. As shown in figure~\ref{fig2_cf}, for a given $\Phi$, the  $c_{f,\Phi}/c_{f,\Phi=0}$ of the oblate case decrease much more rapidly than that of the spherical and prolate cases, indicating that the suspended oblate particles result in a more pronounced shear-thinning effect, which can also be verified by the effective viscosity (see Appendix~\ref{sec:appendixB}). Moreover, for a given $Re$ and $\Phi$, the suspensions of the spherical, prolate and oblate particles result in the greatest, moderate and minimal drag, respectively. In the investigated ranges, the largest discrepancy of $c_{f,\Phi}/c_{f,\Phi=0}$ between suspensions of particles with different aspect ratios is $4\%$, which occurs at $\Phi=10\%$ and $Re=1.3\times10^4$ in the spherical and oblate cases. Given the impressive accuracy of the drag measurements (less than 1\%) and the relatively low $\Phi$, this disparity in the drag modification is quite noticeable. In addition, the difference of $c_{f,\Phi}/c_{f,\Phi=0}$ increases with increasing $\Phi$, which is due to the increasing importance of particle-induced stress at higher $\Phi$ \citep{ardekani2019turbulence}. Therefore,  it is reasonable to speculate that, as $\Phi$ further increases, the differences of the drag modification between different aspect ratios would become larger. Note that, this result is different from the particle-laden channel flow in the previous study \citep{ardekani2019turbulence}, where they found that only the spherical particles increase the drag and the other two types of particles reduce the drag.}

\textcolor{black} {The dependence of the drag on the particle aspect ratio provides an experimental clue to the understanding of the mechanisms of bubbly drag reduction. Flow with dispersed bubbles can, under certain conditions, result in significant drag reduction \citep{van2005drag, van2007bubbly, ezeta2019drag}. However, bubbles are deformable, making it impossible to fix the bubble shape and therefore to isolate the effects of the bubble shape. \cite{van2013importance} and \cite{verschoof2016bubble} have reported that the bubble deformation is crucial for achieving significant drag reduction, but the contributions of (\romannumeral1) the deformation process and (\romannumeral2) the ultimate bubble shape after the deformation remain unknown.}
\textcolor{black} {Obviously, the rigid particle suspensions are significantly different from bubbly flow, including the slip conditions at the surface, the resistance of the dispersed phase to straining motion, the polydispersity, the presence/absence of contaminants (surfactants), etc, and therefore a one-to-one comparison between these two types of flow is also unrealistic. However, the one that should be emphasized is that, as done in this work, the shape effect of rigid spheroidal finite-size particles on the overall drag provides a relatively effortless implement to study the shape effects of bubbles during the deformation process. Notwithstanding how simplistic the rigid particles are compared to bubbles, these particles (rigid particles and bubbles) behave somehow similar in such as the kinematic motion \citep{mathai2020bubbly}, making the study of the rigid particles a reasonable choice to provide a reference case for bubbly flow studies.}
\textcolor{black}{In this work, the rigid spheroidal particles are used to study their shape effect on drag. Due to the frictional contact between solid surfaces, rigid particles dispersed in the flows result in the drag enhancement \citep{guazzelli2018rheology}, which is contrary to the drag reduction in the bubbly flow. However, the fact that the particle aspect ratio affects the drag modification in the current particle-laden flow, also hints that the bubble shape could affect the drag of the bubbly flow.}
\textcolor{black} {Given that the shape effect on the overall drag is relatively small (compared to the drag modulation amplitude in bubbly flow, which is around $40$\% at $\Phi = 4\%$ \citep{verschoof2016bubble}), one would expect that the bubbles can also modulate the turbulence through other mechanisms next to the change of their shapes. Therefore, the unique features of bubbly flow relative to rigid particle-laden flow are worthy of attention for future studies.}
\subsection{The collective effects of particles}
\begin{figure}
	\centering
	\includegraphics[width=1\linewidth]{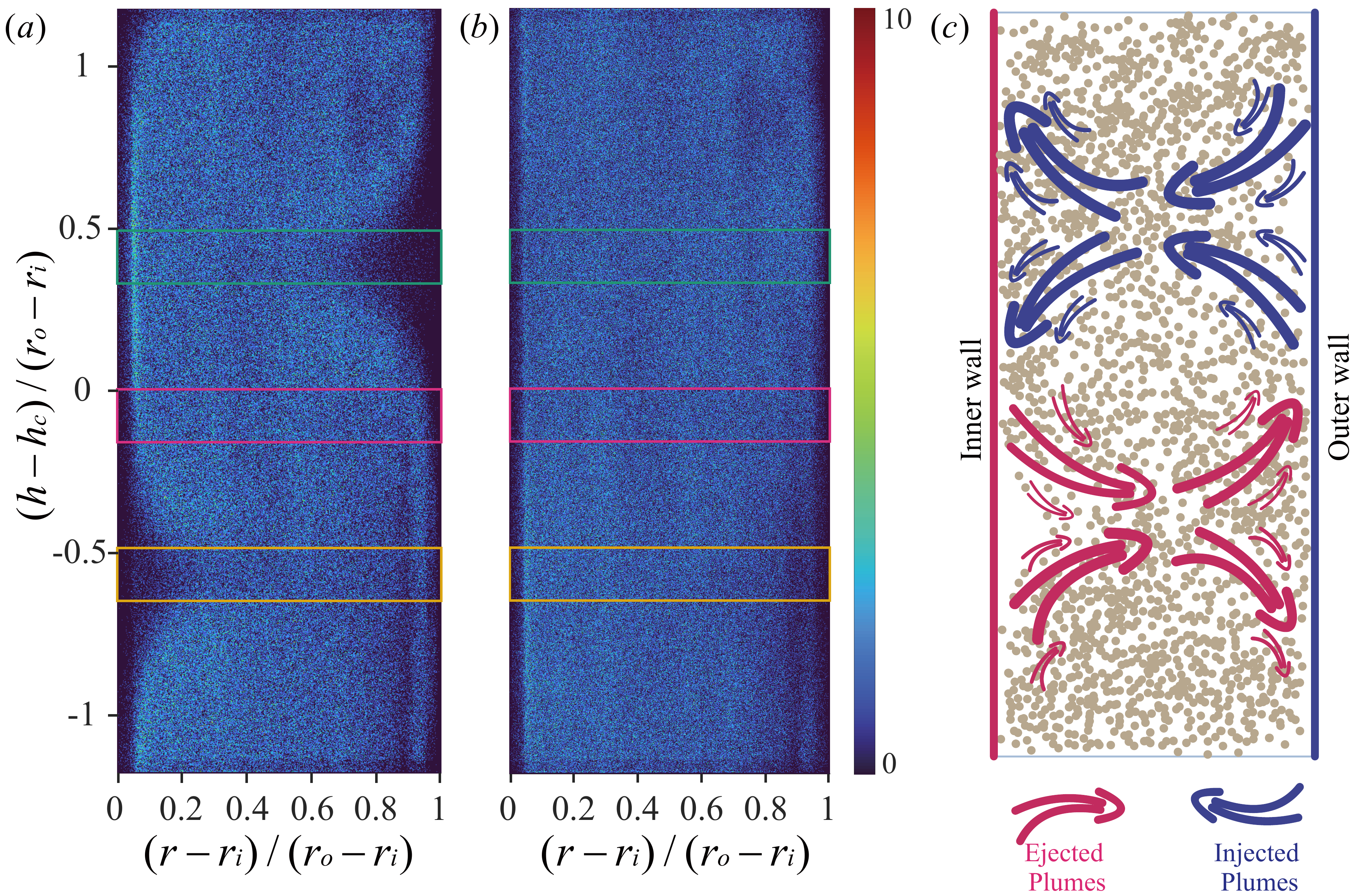}
	\vspace{-5 mm}
	\caption{\label{fig4_taylorroll} \textcolor{black} {The spherical particle positions which are obtained from different frames for $\Phi=0.5\%$ at (a) $Re=6.5\times10^3$  and (b) $Re=1.3\times10^4$. The value of the color bar means the number of times that the particles appear at the position denoted by the points in the graph. The green, magenta and yellow rectangles represent the injected region, vortex-center region and the ejected region, respectively, and the width of which is around $1.7d_e$. Note that most of the points in the graph are corresponding to values of $1\sim2$ in the color bar.} 
	\textcolor{black} { $(c)$ shows the sketch of flow structures and the particle positions (dots) at $Re=6.5\times10^3$. The injected and ejected plumes are represented by the red and blue arrows, respectively. Note that the particle positions in the sketch are not experimental data and for demonstration only.}}
\end{figure}

\textcolor{black}{To reveal the physical mechanism of the drag modification, we look into the particle distribution by performing optical measurements. It should be noted that, even at the minimal particle volume fraction ($\Phi=2\%$), the particles cause violent light scattering and cannot be accurately detected by algorithms.}
\textcolor{black}{Therefore, we start with a smaller volume fraction ($\Phi=0.5\%$, the number of particles $\simeq1700$) to reduce the intensity of the light scattering, so that the particles in the images can be detected using the ellipse detection algorithms \citep{particledetection}. One may question that whether the particle distribution obtained at this low volume fraction could qualitatively represent that at higher volume fractions since the particle dynamics change greatly with increasing $\Phi$. To assuage this misgiving, we perform torque measurements at $\Phi=0.5\%$ and the results have been shown in figures~\ref{fig2_cf},\ref{fig3_effectivevis},\ref{fig8_GRe},\ref{fig9_cf_Rem},\ref{fig10_shearthin}. It can be seen that the quantities related to the drag of the system at $\Phi=0.5\%$ (including $c_f$, $\nu_{eff}$ and $G$) show similar trends to that at higher volume fractions, hinting that the particles behave in similar ways at the low and high volume fractions in the current parameter regimes.}
\textcolor{black}{Here, as depicted in figure~\ref{fig4_taylorroll}(a,c), we define the region where the plumes are ejected from the inner (outer) boundary layer to the bulk as the ejected (injected) region, and the region between them is defined as the vortex-center region. The spherical particle positions which are obtained from different frames when $\Phi=0.5\%$ at $Re=6.5\times10^3$ and $Re=1.3\times10^4$ are given in figure~\ref{fig4_taylorroll}(a) and figure~\ref{fig4_taylorroll}(b), respectively. When $Re=6.5\times10^3$, it is found that the particles distribute non-uniformly, and the pattern of the particle distributions reminisces about the Taylor vortices in TC flow \citep{grossmann2016high}. The particle distributions show a lower clustering near the inner wall in the ejected region, while in the injected region, the lower clustering emerges near the outer wall. However, when the $Re$ is increased to $1.3\times10^4$, the particle distribute nearly homogeneously in the entire system (figure~\ref{fig4_taylorroll}(b)).}

\begin{figure}
	\centering
	\includegraphics[width=1\linewidth]{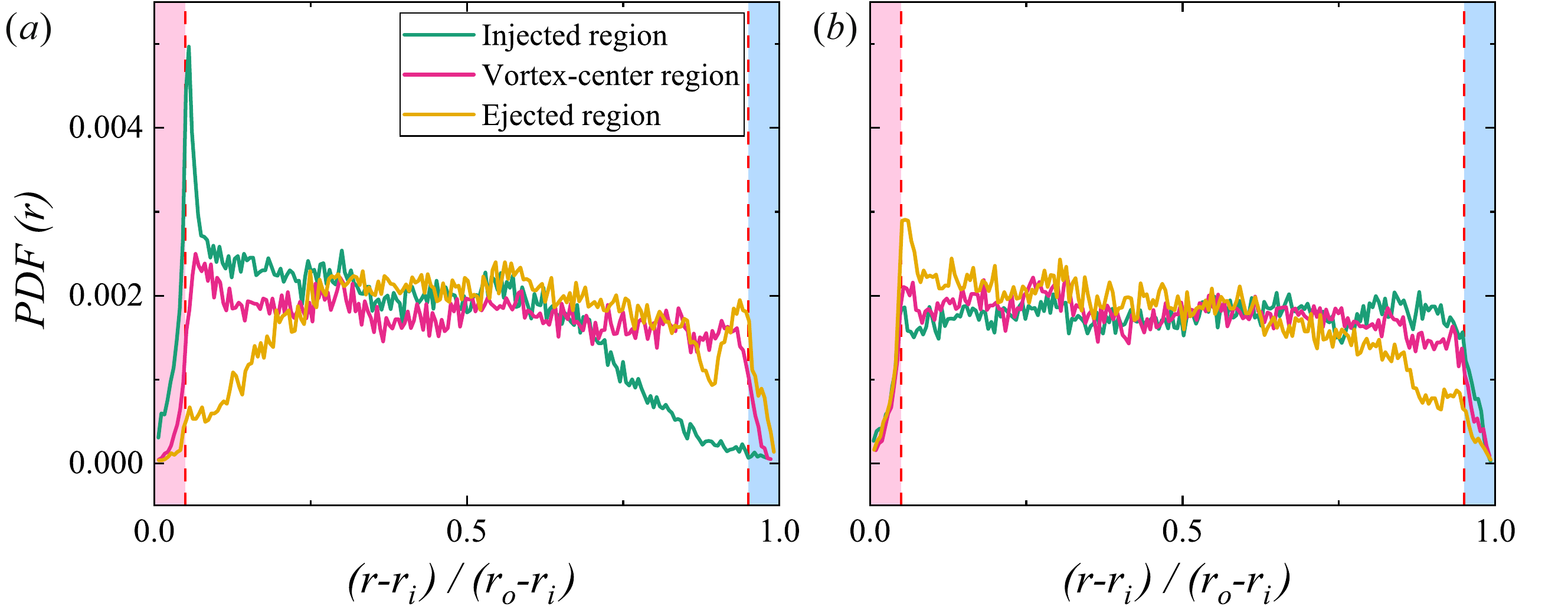}
	\vspace{-5 mm}
	\caption{\label{fig5_PDFspherical} \textcolor{black} {The PDF of the particles position for
		$\lambda=1$ (spherical) at $(a)$ $Re=6.5\times10^3$  and $(b)$ $Re=1.3\times10^4$. The two red dashed vertical lines in each figure indicate the position of an equivalent radius of the particles (i.e., $r_e=0.5d_e$) away from the walls.}}
\end{figure}

\textcolor{black}{To quantitatively evaluate the collective effects of the particles, we choose three typical regions of the flow structures (i.e., the injected region, ejected region, and the vortex-center region) and calculated the PDF of the particle radial positions. As shown in figure~\ref{fig5_PDFspherical}, the PDF is consistent with the pattern of the particle distributions in figure~\ref{fig4_taylorroll}. When $Re=6.5\times10^3$, the PDF has the minimal value near the inner wall in the ejected region, indicating the emergence of the lower clustering region. While in the injected region, the PDF peaks near the inner wall (i.e., highly clustering region) and approaches the lowest value near the outer wall. Not surprisingly, as shown in figure~\ref{fig5_PDFspherical}(b), the PDF curves tend to overlap with each other when the $Re$ increases,}
\textcolor{black} {indicating that the distribution of the particles becomes less non-uniform in the entire system at higher $Re$. Note that, the inner wall peak appearing in the ejected region at higher $Re$ (figure 5b), which is opposite to the behavior found at low $Re$ (figure 5a), and this might result from the slight shift of flow structure positions since the positions of the Taylor vortex could be not exactly the same at low and high $Re$.}

\textcolor{black} {The difference of the particle distributions at different $Re$ can be interpreted from the perspective of the evolution of flow structures and the finite-size effect of the particles. On the one side, though the TC flow displays flow structure at high $Re$ \citep{huisman2014multiple}, the Taylor vortices become turbulent and the velocity fluctuations increase as $Re$ increasing \citep{grossmann2016high}, resulting in the more vigorous particle motion. On the other side, the finite-size effect of the particle can be characterized by the particle Reynolds number.}
\textcolor{black} {Since the slip velocity of the particles is not accessible in the current experiments, the particle Reynolds number is estimated using the bulk velocity of the flow $u_b$, i.e., $Re_p=u_bd_e/\nu$, where $u_b$ can be estimated as $u_b\simeq0.4\cdot \omega_ir_i$ in the present Reynolds number regime \citep{grossmann2016high}.}
\textcolor{black} {Considering that the flow Reynolds number is defined by equation~\ref{Re}, one can obtain the particle Reynolds number as: 
\begin{equation}\label{Rep}
	Re_p=0.4{\frac{d_e}{d}}Re.
\end{equation}
Hence, the $Re_p$ ranges from $2.6\times10^2$ to $10^3$ in the experiments, and the finite-size effects of the particle become more pronounced at higher $Re$.}

\textcolor{black} {Further, the particle inertia can also be measured by the particle Stokes number, $St=\tau_p/\tau_\eta$, where $\tau_p$ and $\tau_\eta$ are the inertia response time of the particles and the Kolmogorov timescale of the flow, respectively. For spherical particles suspended in TC flow, the Stokes number is reduced to: 
\begin{equation}\label{St}
	St={\frac{\rho_pd_p^2}{18\rho_f[(r_o^2-r_i^2)r_id]^{1/2}}(GRe)^{1/2}},
\end{equation}
i.e., the $St$ (or, the particle inertia) increases with increasing $G$ and $Re$. For non-spherical particles, $St_{non}=c\cdot St$, where $c=f(\lambda)$ is a constant only depends on $\lambda$ \citep{voth2017anisotropic}. Since the suspension is dilute here ($\Phi=0.5\%$), the non-dimensional torque $G$ can be approximated to that of the single-phase case, $G_{sp}$, giving the $St$ ranging from $10$ to $60$. Therefore, in the current $Re_p$ and $St$ ranges, the suspended particles could induce unsteady wakes and exhibit inertia clustering \citep{toschi2009lagrangian, mathai2020bubbly}. Specifically, for the case of $Re=6.5\times10^3$ in figure~\ref{fig4_taylorroll}(a) (correspondingly, $Re_p=2.6\times10^2$ and $St=10$), the particles are principally driven by the strong plumes released from the boundary layer, which are depicted by the arrows in figure~\ref{fig4_taylorroll}(c). While for the case of $Re=1.3\times10^4$ in figure~\ref{fig4_taylorroll}(b) (correspondingly, $Re_p=5.2\times10^2$ and $St=25$), the particles with the increased inertia can escape more easily from the flow structures and distribute more uniformly.}

\begin{figure}
	\centering
	\includegraphics[width=1\linewidth]{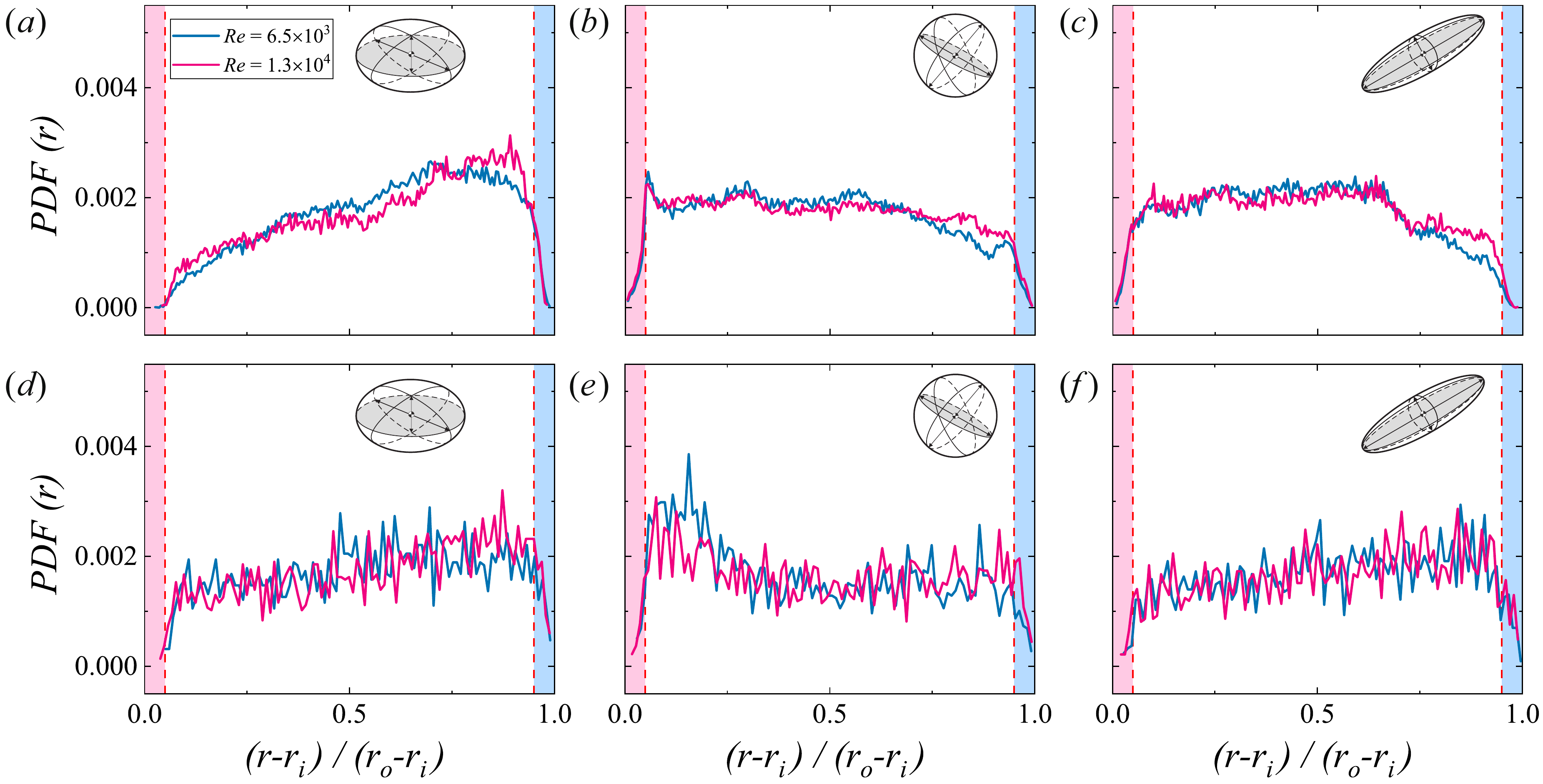}
	\vspace{-5 mm}
	\caption{\label{fig6_pdf_Aver} \textcolor{black} {The averaged PDF of the particles position (a,d) for $\lambda=1/3$ (oblate), (b,e) for $\lambda=1$ (spherical), (c,f) for $\lambda=3$ (prolate). (a-c) are obtained at $\Phi=0.5\%$ by calculating the arithmetic average value of the three typical regions in figure~\ref{fig4_taylorroll},}
	\textcolor{black} {while (d-f) are obtained at $\Phi=2\%$ by manually detecting the particle positions.}
	\textcolor{black} { The two red dashed vertical lines in each figure are the same as in figure~\ref{fig5_PDFspherical}.}}
\end{figure}

\textcolor{black}{Since the particle distributions are inhomogeneous in the axial direction of the system, we use the averaged PDF, which is obtained by calculating the arithmetic average value of the three typical regions, to quantitatively represent the collective effects of particles in the entire system. As shown in figure~\ref{fig6_pdf_Aver}(a-c), for a given $\lambda$, the trends of the averaged PDF curves remain when the $Re$ increases from $6.4\times10^3$ to $Re=1.3\times10^4$, suggesting that the increasing turbulent intensity has negligible effects on the averaged particle migration in the radial direction of the system in the current situation.}

\textcolor{black} {Nevertheless, for different $\lambda$, the averaged PDF curves show distinct differences regardless of the $Re$. In other words, particles with different $\lambda$ show different collective behaviors near the walls and thereby are expected to affect the boundary layer to different degrees. Indeed, the formation of particle layers has been reported in the particle-laden channel flow in the previous simulation works \citep{costa2016universal, ardekani2017drag, ardekani2019turbulence}. \cite{ardekani2019turbulence} have shown that the spherical particles form a particle layer, which could enhance the near-wall Reynolds stress. While for the cases of the oblate and prolate particles, the near-wall Reynolds stress is attenuated due to the absence of the particle layer. In the present work in TC flow, as indicated in figure~\ref{fig5_PDFspherical} and figure~\ref{fig6_pdf_Aver}(b), the spherical particles preferentially cluster and form a particle layer near the inner wall, where the boundary layer exists. Therefore, the maximum drag modification is observed in the flow laden with spherical particles. However, the near-wall clustering phenomenon disappears in the oblate case (figure~\ref{fig6_pdf_Aver}(a)), of which the PDF peaks in the bulk and has the lowest value near the walls. The averaged PDF curves indicate that the oblate particles prefer to cluster in the bulk and thereby have a smaller effect on the boundary layer, accounting for the observations of the minimal drag modification in figure~\ref{fig2_cf} and figure~\ref{fig3_effectivevis}.}

\textcolor{black} {However, one more question remains since so far the analyses of particle distribution are based on the optical measurements at $\Phi=0.5\%$: will the particle preferential clustering persist at even higher $\Phi$? To bridge the gap of volume fractions between the torque measurements ($\Phi\geq2\%$) and optical measurements ($\Phi=0.5\%$), we further conduct optical measurements at higher volume fractions. Since the domain is illuminated from the back, the light intensity decreases greatly at higher volume fractions, making the particles hard to be distinguished from the background fluid. At higher volume fractions, the feasible method to detect the particles is manual detection, which is of relatively low accuracy but provides reliable information. The averaged PDF curves of particle distribution at $\Phi=2\%$ are shown in figure~\ref{fig6_pdf_Aver}(d-f). For each case, 100 frames are used and the total number of detected particles is more than 3000. For even higher volume fractions ($\Phi\geq4\%$), the enormous amount of unfocused particles make the particles inside the focal plane invisible, therefore no useful data can be obtained.}

\textcolor{black} {One can see in figure~\ref{fig6_pdf_Aver}(d-f) that the spherical particles still preferentially cluster near the inner wall and therefore form a particle layer, which is consistent with the result obtained at $\Phi=0.5\%$. Additionally, at $\Phi=2\%$, the distribution of spherical particles becomes more non-uniform in the radial direction than that at $\Phi=0.5\%$, which has also been reported in \cite{fornari2016effect}, hinting that the spherical particles tend to move toward the walls as $\Phi$ increases. This growing non-uniform distribution possibly results from the stronger shear-induced particle-particle interactions at higher $\Phi$, and could partially account for the increasing drag differences compared with other particle shapes as $\Phi$ increases. Moreover, as $\Phi$ increases, the particles might relaminarize the bulk flow, and thus suppress the Reynolds stress \citep{fornari2016effect}. This Reynolds stress suppression could therefore yield a stronger $Re$-dependence of the drag at higher $\Phi$, which is verified by the stronger shear-thinning effects of normalized friction coefficient (figure \ref{fig2_cf}) and the normalized effective viscosity (figure \ref{fig10_shearthin}). On the other hand, for oblate and prolate cases, it is clear that most of the particles distribute in the bulk region. The near-wall clustering phenomenon disappears and therefore the particle layer is absent, which is similar to that of $\Phi=0.5\%$ and accounts for their similar overall drag.}

\textcolor{black} {So far, the optical measurements performed at $\Phi=0.5\%$ and $\Phi=2\%$ lead to the same conclusion that the larger drag modification is connected to the near-wall particle clustering in the current system. Though the result and conclusion of particle preferential clustering might not be directly extrapolated further to higher volume fractions, one reasonable speculation would be that it will play a significant role in the drag modulation at higher $\Phi$. Of course, further studies are needed to confirm it in future work.}

\section{Conclusion}\label{sec:conclusion}
\textcolor{black} {In this study, we experimentally investigated the drag modification by neutrally buoyant finite-size particles in the Taylor-Couette turbulent flow. The Reynolds number ranges from $6.5\times10^3$ to $2.6\times10^4$, and the particle volume fraction ranges from $0\%$ to $10\%$. To study the shape effects of the finite-size particle, we conduct experiments using particles with aspect ratio equals to $\lambda=1/3$ (oblate), $\lambda=1$ (spherical), and $\lambda=3$ (prolate). It is found that, different from the cases in bubbly TC flow, the rigid particles increase the drag of the system due to the frictional contact between solid surfaces, and the particle-laden flow somehow exhibits the shear-thinning effect since the turbulent stress becomes dominant at higher Reynolds number in the current low volume fraction ranges. The drag modification by the particles was also interpreted from the perspective of the effective viscosity, which shows a discrepancy from the Krieger \& Dougherty formula due to the non-negligible flow inertia and the non-uniform particle distributions at high $Re$ and $\Phi$ in the current study. Through varying the particle aspect ratios, we found that the suspensions of the spherical, prolate, and oblate particles result in the greatest, moderate and minimal drag, respectively. The dependence of the drag modification on the particle aspect ratio also hints that the bubble shape might affect the drag of the flow to some extent in bubbly flow.}

\textcolor{black} {Furthermore, we performed optical measurements}
\textcolor{black} {at low volume fractions ($\Phi=0.5\%$ and $\Phi=2\%$)}
\textcolor{black} { and found that the drag modification induced by the suspended particles is related to the particle collective effects. For $Re=6.5\times10^3$ (correspondingly, $Re_p=2.6\times10^2$ and $St=10$) at $\Phi=0.5\%$, the particles could follow the motion of the flow and the distributions of the particle positions obtained from different frames reflect the Taylor vortices. However, when $Re$ increases to $1.3\times10^4$ (correspondingly, $Re_p=5.2\times10^2$ and $St=25$), the Taylor vortices become turbulent and the finite-size effects of the particle increase, allowing the particles to escape more easily from the flow structures and resulting in the more uniform distributions. Additionally, it is found that the particle aspect ratio affects the particle collective effects. The spherical particles preferentially cluster near the walls and form a particle layer, while which are absent in the oblate case.}
\textcolor{black} {This particle preferential clustering is verified to hold at $\Phi=2\%$. Based on the results, it can be concluded that, for particles with different aspect ratios at low volume fractions ($\Phi=0.5\%$ and $\Phi=2\%$ here), the different preferential collective effects could affect the boundary layer to different degrees, and finally result in the different drag modifications. More studies are needed to further explore how particles modify the overall drag at higher volume fractions.}

\section*{Acknowledgement}
\textcolor{black} {This work was supported by the Natural Science Foundation of China under grant nos.11988102, 91852202.}

\appendix 
\section{Torque measurement and the end effect calibration}\label{sec:appendixA}
\begin{figure}
	\centering
	\includegraphics[width=0.5\linewidth]{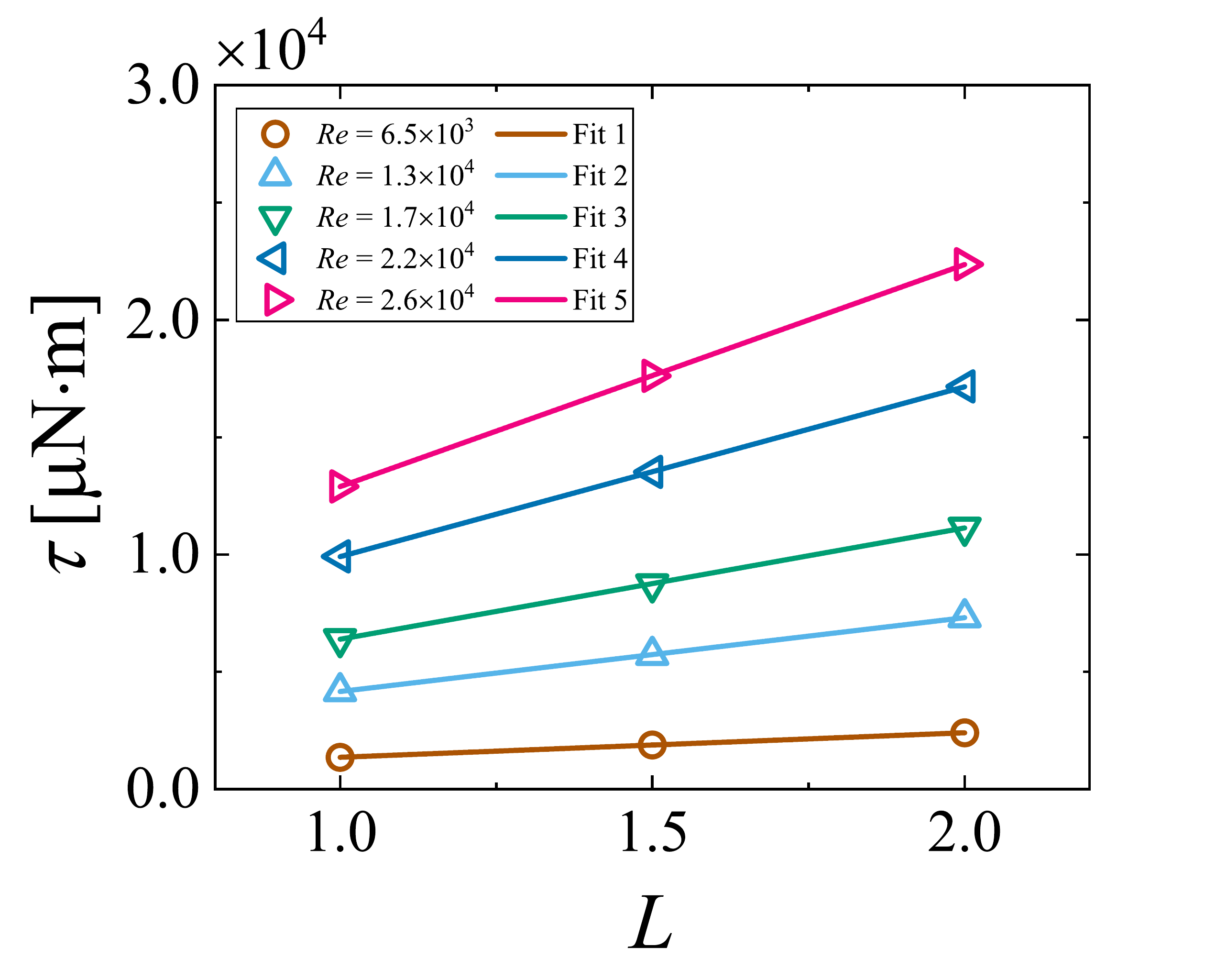}
	\vspace{-2 mm}
	\caption{\label{fig7_ending_effect} The calibration of the torque measurements. The torque contribution from the von K\'{a}rm\'{a}n flow at the ends of the system can be determined as the vertical intercept of the fitting line. Here the particle volume fraction $\Phi=0$, i.e., the single-phase glycerin-water solution, is used. }
\end{figure}

{{ 
		The torque of system, $\tau_{total}$, can be divided into two parts: (\romannumeral1) the torque, $\tau$, due to the sidewall of the inner cylinder (TC flow), (\romannumeral2) the torque, $\tau_{end}$, contributed from the von K\'{a}rm\'{a}n flow at the end plates. Since $\tau$ increases linearly with the height of the inner cylinder \citep{greidanus2011drag, hu2017significant}, the $\tau_{end}$ can be obtained through the vertical intercept of the fitting lines. We conduct experiments to measure the torque using three apparatus with different heights of $L$, $1.5L$, $2L$, and the results are shown in figure~\ref{fig7_ending_effect}. In these calibration experiments, the single-phase glycerin-water solution is used (i.e., $\Phi=0$). For different Reynolds numbers, the ratio of $\tau_{end}$ to $\tau_{total}$ is found to be around a constant (25\%). Here we assume that for $\Phi>0$ the ratio of $\tau_{end}$ to $\tau_{total}$ is the same as that for single-phase case ($\Phi=0$). }}
	
\begin{figure}
	\centering
	\includegraphics[width=1\linewidth]{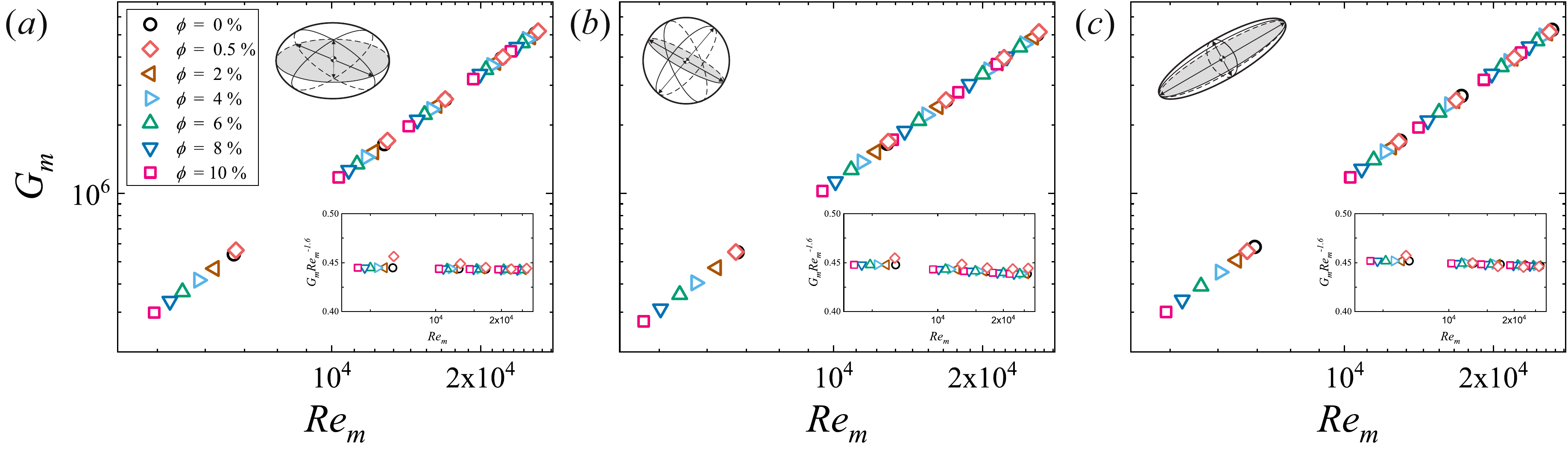}
	\vspace{-5 mm}
	\caption{\label{fig8_GRe} The dependence of the modified dimensionless torque, $G_m$, on the modified Reynolds number, $Re_m$, for (a) $\lambda=1/3$ (oblate), (b) $\lambda=1$ (spherical), (c) $\lambda=3$ (prolate). The insets show $G_m$ compensated with $Re_m^{-1.6}$.}
\end{figure}

\section{The calculation of the effective viscosity}\label{sec:appendixB}
\begin{figure}
	\centering
	\includegraphics[width=1\linewidth]{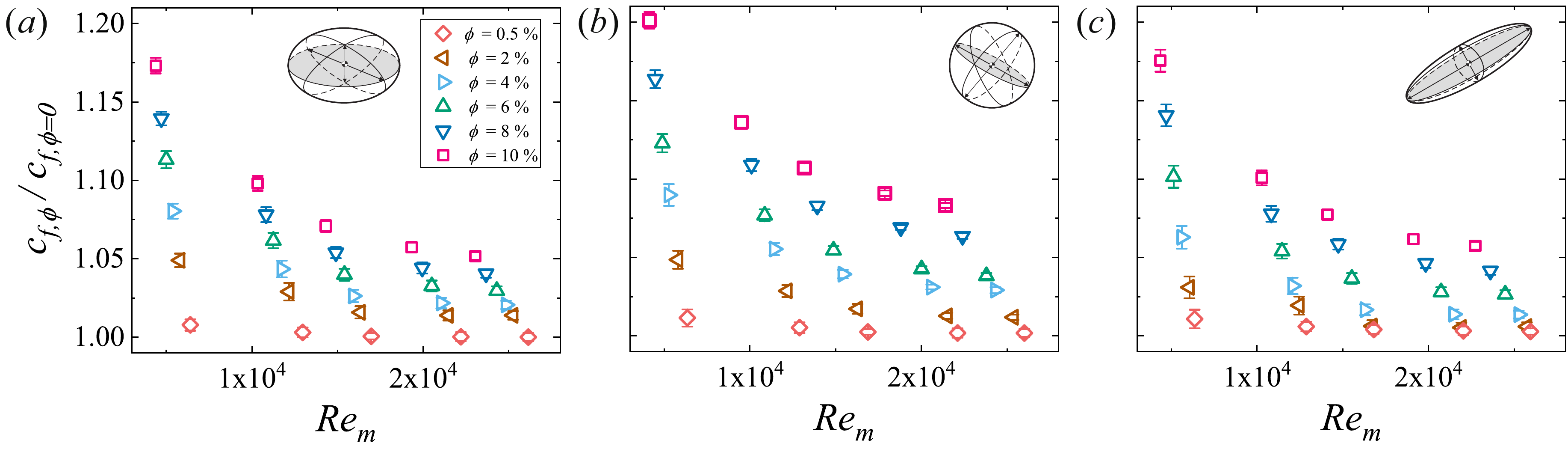}
	\vspace{-5 mm}
	\caption{\label{fig9_cf_Rem} \textcolor{black} {The normalized friction coefficient of the Taylor-Couette system, $c_{f,\Phi}/c_{f,\Phi=0}$, as a function of the modified Reynolds number, $Re_m$, for (a) $\lambda=1/3$ (oblate), (b) $\lambda=1$ (spherical) and (c) $\lambda=3$ (prolate), where $c_{f,\Phi=0}$ is the friction coefficient for the single-phase case. The accuracy of the experiments is indicated by the error bars, which are typically less than 1\%.}}
\end{figure}

\textcolor{black} {The rheology of multiphase flow can be characterized using the effective viscosity, $\nu_{eff}$. For single-phase TC turbulent flow ($\Phi=0\%$), a simple scaling law exists between $G$ and $Re$: $G\propto Re^\alpha$ \citep{eckhardt2000scaling}. Assuming that this relation is still valid for particle-laden flow ($\Phi>0\%$) here, the following equation can be derived:
\begin{equation}\label{a1}
{\nu_{eff}\over\nu_f}={({\tau\over{\tau_f}})^{1\over{2-\alpha}}},
\end{equation}
where $\nu_f$ is the viscosity of the fluid (i.e., the single-phase flow), and  $\tau$ and $\tau_f$ are the measured torque for the particle-laden flow and the single-phase flow, respectively.}
\textcolor{black} {As one could see from the derivation above, the effective viscosity is a fitting parameter depending on the scaling law between $G$ and $Re$.}
\textcolor{black} {Using the equation above, the effective viscosity of the particle-laden flow can be calculated. We define the modified dimensionless torque, $G_m={\tau \over{2\pi l\rho_f\nu_{eff}^2}}$, and the modified Reynolds number, $Re_m=\omega_ir_id/\nu_{eff}$. As shown in figure~\ref{fig8_GRe}, it is found that the particle-laden flow satisfied the scaling law $G_m\propto Re_m^\alpha$, where $\alpha\simeq1.6$.}
\textcolor{black} {The dependence of the normalized friction coefficient on the modified Reynolds number is shown in figure \ref{fig9_cf_Rem}.}

\begin{figure}
	\centering
	\includegraphics[width=1\textwidth]{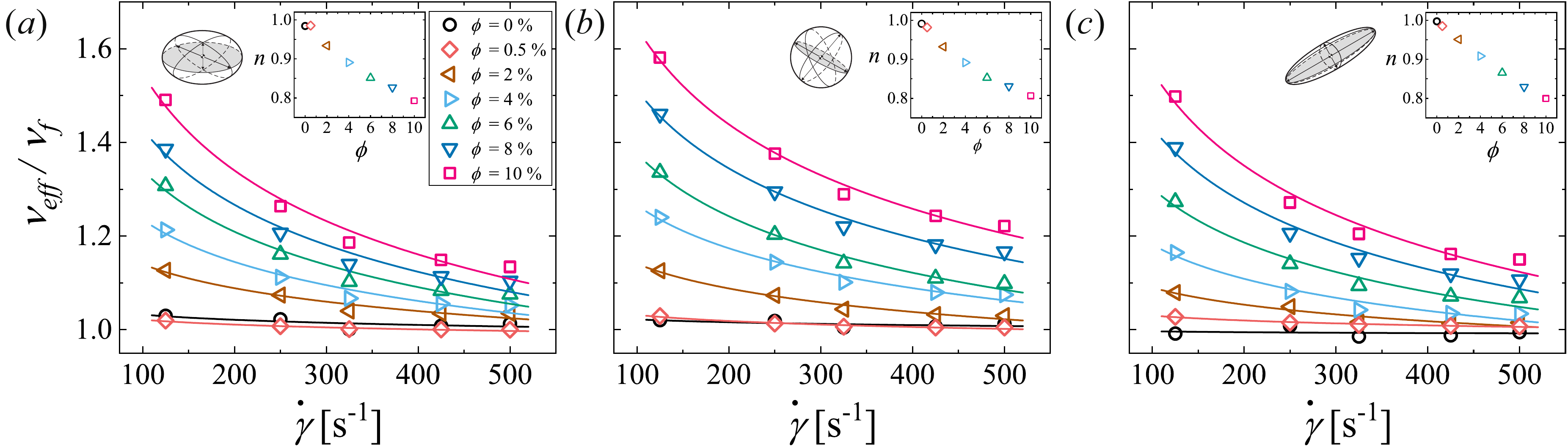}
	\vspace{-5 mm}
	\caption{\label{fig10_shearthin} The dependence of the non-dimensional effective viscosity, $\nu_{eff}/\nu_f$, on the characteristic shearing rate, $\dot{\gamma}$, for (a) $\lambda=1/3$ (oblate), (b) $\lambda=1$ (spherical) and (c) $\lambda=3$ (prolate). The curves are fitting results using the HB model. Insets: the decreasing flow index $n$ with the increasing $\Phi$ indicates a more pronounced shear-thinning effect at higher $\Phi$.}
\end{figure} 
\textcolor{black} {Using the effective viscosity, the shear-thinning effect of the particle-laden flow can be quantified. The dependence of the non-dimensional effective viscosity, $\nu_{eff}/\nu_f$, on the characteristic shearing rate, $\dot{\gamma}=\omega_ir_i/d$, for various aspect ratios is shown in figure~\ref{fig10_shearthin}. To evaluate the shear-thinning effect of the particle-laden flow, we use the Herschel-Bulkley (HB) model \citep{herschel1926konsistenzmessungen}:
\begin{equation}\label{HB}
	\mu_{eff}=k_0\dot{\gamma}^{n-1}+\tau_0\dot{\gamma}^{-1},
\end{equation}
where $\mu_{eff}$ is the effective dynamic viscosity of the particle-laden flow, $k_0$ and $n$ represents the consistency and the flow index, respectively, and $\tau_0$ is the yield shear stress. Since the flow is far from the jamming state in the current system, the yield shear stress is expected to be zero ($\tau_0=0$).
As the density of the fluid and the particle are matched, we take $\rho_f$ as the density of the particle-laden flow, giving $\mu_{eff}/\mu_f$ = $\nu_{eff}/\nu_f$, where $\mu_f$ is the dynamic viscosity of the single-phase flow. Consequently, the HB model can be simplified as $\nu_{eff}/{\nu_f}=k_0\dot{\gamma}^{n-1}/\mu_f$ and the fitting curves are shown in figure~\ref{fig10_shearthin}. Note that the fitting parameter $k_0$ is different for various particle aspect ratios and volume fractions. For the single-phase case ($\Phi=0\%$), the normalized effective viscosity $\nu_{eff}/\nu_f$ is around 1 for various $Re$. When the $\Phi$ increases, the HB model agrees well with the experimental data, and the flow index $n$ of the HB model decreases with the increasing $\Phi$, suggesting that the shear-thinning effect becomes more pronounced at higher $\Phi$.}

\begin{figure}
	\centering
	\includegraphics[width=0.9\linewidth]{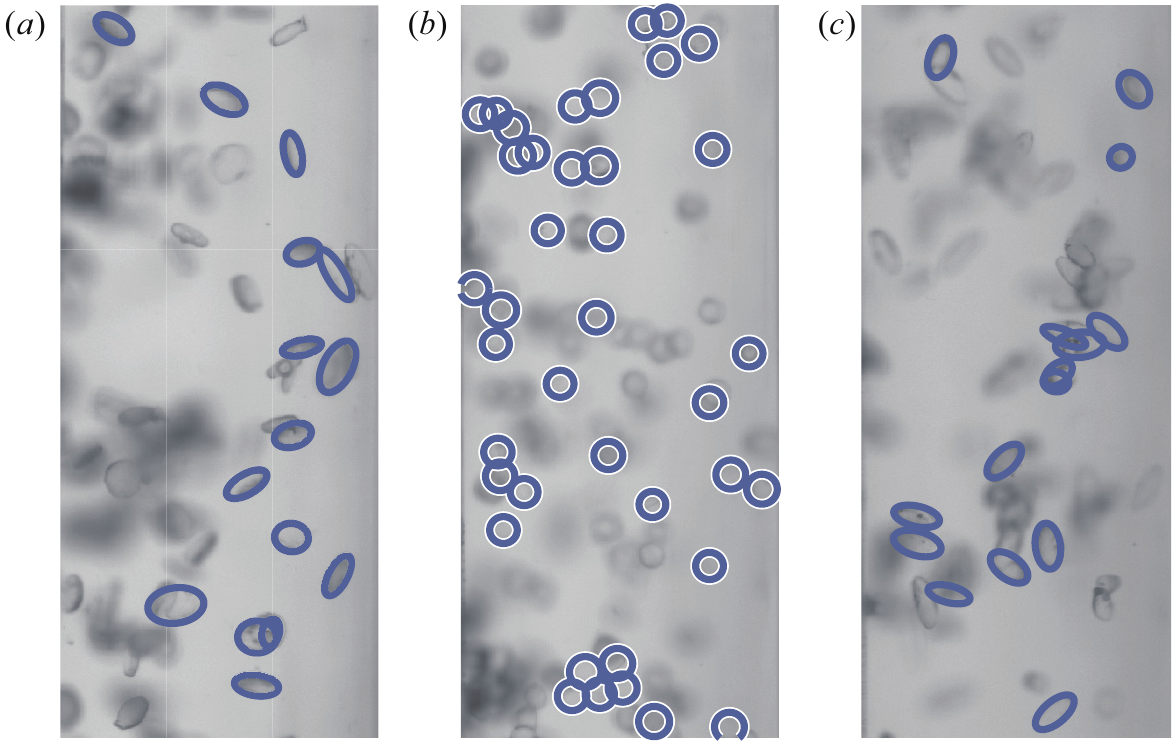}
	\vspace{-1 mm}
	\caption{\label{fig11_detection} The examples of particle detection for (a) $\lambda=1/3$ (oblate), (b) $\lambda=1$ (spherical) and (c) $\lambda=3$ (prolate) at $Re=6.5\times10^3$. Most particles inside the focal plane have been detected.}
\end{figure}
\section{Particle detection}\label{sec:appendixC}
\textcolor{black} {During the optical measurements, we locate the focal plane of the camera at the crossing plane of the axis and the light-line (the cyan shadow plane in figure~\ref{fig1_sketch}(a)). The depth of the field is around $0.21$\ mm, which is smaller than the particle diameter. Therefore the depth resolution in the azimuthal direction is high enough. The central height of the camera is kept the same as the apparatus so that the visible window is located at the central height of the apparatus, therefore the ending effects in the axis direction are eliminated. }

\textcolor{black} {The particle detection is performed using the ellipse detection method proposed by \cite{particledetection}. The examples of particle detection are given in figure~\ref{fig11_detection}. Inside the focal plane, most particles are detected and a few others are not since the material used to print the particles is transparent. Therefore, to ensure the accuracy of the data, 30000 frames are used to calculate the cumulative spatial distribution (figure.~\ref{fig4_taylorroll}) and the PDF curves (figure.~\ref{fig5_PDFspherical} and figure.~\ref{fig6_pdf_Aver}) for each $Re$ case.}


\begin{thebibliography}{66}
\expandafter\ifx\csname natexlab\endcsname\relax\def\natexlab#1{#1}\fi
\def\au#1{#1} \def\ed#1{#1} \def\yr#1{#1}\def\at#1{#1}\def\jt#1{\textit{#1}}
  \def\bt#1{#1}\def\bvol#1{\textbf{#1}} \def\vol#1{#1} \def\pg#1{#1}
  \def\publ#1{#1}\def\arxiv#1{#1}\def\org#1{#1}\def\st#1{\textit{#1}}

\bibitem[Adams {\em et~al.\/}(2004)Adams, Frith \& Stokes]{adams2004influence}
{\sc \au{Adams, S}, \au{Frith, WJ} \& \au{Stokes, JR}} \yr{2004}  \at{Influence
  of particle modulus on the rheological properties of agar microgel
  suspensions}.  \jt{J. Rheol.}  \bvol{48}~(6),  \pg{1195--1213}.

\bibitem[Alm{\'e}ras {\em et~al.\/}(2017)Alm{\'e}ras, Mathai, Lohse \&
  Sun]{almeras2017experimental}
{\sc \au{Alm{\'e}ras, Elise}, \au{Mathai, Varghese}, \au{Lohse, Detlef} \&
  \au{Sun, Chao}} \yr{2017}  \at{Experimental investigation of the turbulence
  induced by a bubble swarm rising within incident turbulence}.  \jt{J. Fluid
  Mech.}  \bvol{825},  \pg{1091--1112}.

\bibitem[Ardekani \& Brandt(2019)]{ardekani2019turbulence}
{\sc \au{Ardekani, M~Niazi} \& \au{Brandt, Luca}} \yr{2019}  \at{Turbulence
  modulation in channel flow of finite-size spheroidal particles}.  \jt{J.
  Fluid Mech.}  \bvol{859},  \pg{887--901}.

\bibitem[Ardekani {\em et~al.\/}(2017)Ardekani, Costa, Breugem, Picano \&
  Brandt]{ardekani2017drag}
{\sc \au{Ardekani, M~Niazi}, \au{Costa, Pedro}, \au{Breugem, W-P}, \au{Picano,
  Francesco} \& \au{Brandt, Luca}} \yr{2017}  \at{Drag reduction in turbulent
  channel flow laden with finite-size oblate spheroids}.  \jt{J. Fluid Mech.}
  \bvol{816},  \pg{43--70}.

\bibitem[Assen {\em et~al.\/}(2021)Assen, Ng, Will, Stevens, Lohse \&
  Verzicco]{assen2021strong}
{\sc \au{Assen, Martin}, \au{Ng, Chong~Shen}, \au{Will, Jelle~B}, \au{Stevens,
  Richard~JAM}, \au{Lohse, Detlef} \& \au{Verzicco, Roberto}} \yr{2021}
  \at{Strong alignment of prolate ellipsoids in taylor-couette flow}.
  \jt{arXiv preprint arXiv:2106.05603} .

\bibitem[Bagnold(1954)]{bagnold1954experiments}
{\sc \au{Bagnold, Ralph~Alger}} \yr{1954}  \at{Experiments on a gravity-free
  dispersion of large solid spheres in a newtonian fluid under shear}.
  \jt{Proc. R. Soc. Lond. A}  \bvol{225}~(1160),  \pg{49--63}.

\bibitem[Bakhuis {\em et~al.\/}(2021)Bakhuis, Ezeta, Bullee, Marin, Lohse, Sun
  \& Huisman]{bakhuis2021catastrophic}
{\sc \au{Bakhuis, Dennis}, \au{Ezeta, Rodrigo}, \au{Bullee, Pim~A}, \au{Marin,
  Alvaro}, \au{Lohse, Detlef}, \au{Sun, Chao} \& \au{Huisman, Sander~G}}
  \yr{2021}  \at{Catastrophic phase inversion in high-reynolds-number turbulent
  taylor-couette flow}.  \jt{Phys. Rev. Lett.}  \bvol{126}~(6),  \pg{064501}.

\bibitem[Bakhuis {\em et~al.\/}(2018)Bakhuis, Verschoof, Mathai, Huisman, Lohse
  \& Sun]{bakhuis2018finite}
{\sc \au{Bakhuis, Dennis}, \au{Verschoof, Ruben~A}, \au{Mathai, Varghese},
  \au{Huisman, Sander~G}, \au{Lohse, Detlef} \& \au{Sun, Chao}} \yr{2018}
  \at{Finite-sized rigid spheres in turbulent taylor--couette flow: effect on
  the overall drag}.  \jt{J. Fluid Mech.}  \bvol{850},  \pg{246--261}.

\bibitem[Batchelor(1970)]{batchelor1970stress}
{\sc \au{Batchelor, GK}} \yr{1970}  \at{The stress system in a suspension of
  force-free particles}.  \jt{J. Fluid Mech.}  \bvol{41}~(3),  \pg{545--570}.

\bibitem[van~den Berg {\em et~al.\/}(2007)van~den Berg, van Gils, Lathrop \&
  Lohse]{van2007bubbly}
{\sc \au{van~den Berg, Thomas~H}, \au{van Gils, Dennis~PM}, \au{Lathrop,
  Daniel~P} \& \au{Lohse, Detlef}} \yr{2007}  \at{Bubbly turbulent drag
  reduction is a boundary layer effect}.  \jt{Phys. Rev. Lett.}  \bvol{98}~(8),
   \pg{084501}.

\bibitem[Van~den Berg {\em et~al.\/}(2005)Van~den Berg, Luther, Lathrop \&
  Lohse]{van2005drag}
{\sc \au{Van~den Berg, Thomas~H}, \au{Luther, Stefan}, \au{Lathrop, Daniel~P}
  \& \au{Lohse, Detlef}} \yr{2005}  \at{Drag reduction in bubbly taylor-couette
  turbulence}.  \jt{Phys. Rev. Lett.}  \bvol{94}~(4),  \pg{044501}.

\bibitem[Calzavarini {\em et~al.\/}(2008)Calzavarini, Cencini, Lohse, Toschi
  {\em et~al.\/}]{calzavarini2008quantifying}
{\sc \au{Calzavarini, Enrico}, \au{Cencini, Massimo}, \au{Lohse, Detlef},
  \au{Toschi, Federico} \& \au{others}} \yr{2008}  \at{Quantifying
  turbulence-induced segregation of inertial particles}.  \jt{Phys. Rev. Lett.}
   \bvol{101}~(8),  \pg{084504}.

\bibitem[Colin {\em et~al.\/}(2012)Colin, Fabre \& Kamp]{colin2012turbulent}
{\sc \au{Colin, Catherine}, \au{Fabre, Jean} \& \au{Kamp, Arjan}} \yr{2012}
  \at{Turbulent bubbly flow in pipe under gravity and microgravity conditions}.
   \jt{J. Fluid Mech.}  \bvol{711},  \pg{469--515}.

\bibitem[Costa {\em et~al.\/}(2016)Costa, Picano, Brandt \&
  Breugem]{costa2016universal}
{\sc \au{Costa, Pedro}, \au{Picano, Francesco}, \au{Brandt, Luca} \&
  \au{Breugem, Wim-Paul}} \yr{2016}  \at{Universal scaling laws for dense
  particle suspensions in turbulent wall-bounded flows}.  \jt{Phys. Rev. Lett.}
   \bvol{117},  \pg{134501}.

\bibitem[Cristancho {\em et~al.\/}(2011)Cristancho, Delgado, Martinez,
  Abolghassemi~Fakhree \& Jouyban]{cristancho2011volumetric}
{\sc \au{Cristancho, Diana~M}, \au{Delgado, Daniel~R}, \au{Martinez, Fleming},
  \au{Abolghassemi~Fakhree, Mohammad~A} \& \au{Jouyban, Abolghasem}} \yr{2011}
  \at{Volumetric properties of glycerol+ water mixtures at several temperatures
  and correlation with the jouyban-acree model}.  \jt{Rev. colomb. cienc. quim.
  farm.}  \bvol{40}~(1),  \pg{92--115}.

\bibitem[Eckhardt {\em et~al.\/}(2000)Eckhardt, Grossmann \&
  Lohse]{eckhardt2000scaling}
{\sc \au{Eckhardt, Bruno}, \au{Grossmann, Siegfried} \& \au{Lohse, Detlef}}
  \yr{2000}  \at{Scaling of global momentum transport in taylor-couette and
  pipe flow}.  \jt{Eur. Phys. J. B}  \bvol{18}~(3),  \pg{541--544}.

\bibitem[Elghobashi(1994)]{elghobashi1994predicting}
{\sc \au{Elghobashi, Said}} \yr{1994}  \at{On predicting particle-laden
  turbulent flows}.  \jt{Appl. Sci. Res.}  \bvol{52}~(4),  \pg{309--329}.

\bibitem[Ezeta {\em et~al.\/}(2019)Ezeta, Bakhuis, Huisman, Sun \&
  Lohse]{ezeta2019drag}
{\sc \au{Ezeta, Rodrigo}, \au{Bakhuis, Dennis}, \au{Huisman, Sander~G},
  \au{Sun, Chao} \& \au{Lohse, Detlef}} \yr{2019}  \at{Drag reduction in
  boiling taylor--couette turbulence}.  \jt{J. Fluid Mech.}  \bvol{881},
  \pg{104--118}.

\bibitem[Fall {\em et~al.\/}(2010)Fall, Lemaitre, Bertrand, Bonn \&
  Ovarlez]{fall2010shear}
{\sc \au{Fall, Abdoulaye}, \au{Lemaitre, Anael}, \au{Bertrand, Fran{\c{c}}ois},
  \au{Bonn, Daniel} \& \au{Ovarlez, Guillaume}} \yr{2010}  \at{Shear thickening
  and migration in granular suspensions}.  \jt{Phys. Rev. Lett.}
  \bvol{105}~(26),  \pg{268303}.

\bibitem[Fornari {\em et~al.\/}(2016)Fornari, Formenti, Picano \&
  Brandt]{fornari2016effect}
{\sc \au{Fornari, Walter}, \au{Formenti, Alberto}, \au{Picano, Francesco} \&
  \au{Brandt, Luca}} \yr{2016}  \at{The effect of particle density in turbulent
  channel flow laden with finite size particles in semi-dilute conditions}.
  \jt{Phys. Fluids}  \bvol{28}~(3),  \pg{033301}.

\bibitem[Frankel {\em et~al.\/}(2016)Frankel, Pouransari, Coletti \&
  Mani]{frankel2016settling}
{\sc \au{Frankel, Ari}, \au{Pouransari, Hadi}, \au{Coletti, Filippo} \&
  \au{Mani, Ali}} \yr{2016}  \at{Settling of heated particles in homogeneous
  turbulence}.  \jt{J. Fluid Mech.}  \bvol{792},  \pg{869--893}.

\bibitem[van Gils {\em et~al.\/}(2013)van Gils, Guzman, Sun \&
  Lohse]{van2013importance}
{\sc \au{van Gils, Dennis~PM}, \au{Guzman, Daniela~Narezo}, \au{Sun, Chao} \&
  \au{Lohse, Detlef}} \yr{2013}  \at{The importance of bubble deformability for
  strong drag reduction in bubbly turbulent taylor--couette flow}.  \jt{J.
  Fluid Mech.}  \bvol{722},  \pg{317--347}.

\bibitem[Greidanus {\em et~al.\/}(2011)Greidanus, Delfos \&
  Westerweel]{greidanus2011drag}
{\sc \au{Greidanus, A~J}, \au{Delfos, R} \& \au{Westerweel, J}} \yr{2011}
  \at{Drag reduction by surface treatment in turbulent taylor-couette flow}.
  \jt{J. Physics: Conf. Ser.}  \bvol{318}~(8),  \pg{082016}.

\bibitem[Grossmann {\em et~al.\/}(2016)Grossmann, Lohse \&
  Sun]{grossmann2016high}
{\sc \au{Grossmann, Siegfried}, \au{Lohse, Detlef} \& \au{Sun, Chao}} \yr{2016}
   \at{High--reynolds number taylor-couette turbulence}.  \jt{Annu. Rev. Fluid
  Mech.}  \bvol{48}.

\bibitem[Guazzelli \& Pouliquen(2018)]{guazzelli2018rheology}
{\sc \au{Guazzelli, {\'E}lisabeth} \& \au{Pouliquen, Olivier}} \yr{2018}
  \at{Rheology of dense granular suspensions}.  \jt{J. Fluid Mech.}
  \bvol{852}.

\bibitem[Herschel \& Bulkley(1926)]{herschel1926konsistenzmessungen}
{\sc \au{Herschel, Winslow~H} \& \au{Bulkley, Ronald}} \yr{1926}
  \at{Konsistenzmessungen von gummi-benzoll{\"o}sungen}.  \jt{Kolloidn. Z.}
  \bvol{39}~(4),  \pg{291--300}.

\bibitem[Hu {\em et~al.\/}(2017)Hu, Wen, Bao, Jia, Song, Song, Pan, Scaraggi,
  Dini, Xue {\em et~al.\/}]{hu2017significant}
{\sc \au{Hu, Haibao}, \au{Wen, Jun}, \au{Bao, Luyao}, \au{Jia, Laibing},
  \au{Song, Dong}, \au{Song, Baowei}, \au{Pan, Guang}, \au{Scaraggi, Michele},
  \au{Dini, Daniele}, \au{Xue, Qunji} \& \au{others}} \yr{2017}
  \at{Significant and stable drag reduction with air rings confined by
  alternated superhydrophobic and hydrophilic strips}.  \jt{Sci. Adv.}
  \bvol{3}~(9),  \pg{e1603288}.

\bibitem[Huisman {\em et~al.\/}(2014)Huisman, Van Der~Veen, Sun \&
  Lohse]{huisman2014multiple}
{\sc \au{Huisman, Sander~G}, \au{Van Der~Veen, Roeland~Ca}, \au{Sun, Chao} \&
  \au{Lohse, Detlef}} \yr{2014}  \at{Multiple states in highly turbulent
  taylor--couette flow}.  \jt{Nat. Commun.}  \bvol{5}~(1),  \pg{1--5}.

\bibitem[Hunt {\em et~al.\/}(2002)Hunt, Zenit, Campbell \&
  Brennen]{hunt2002revisiting}
{\sc \au{Hunt, ML}, \au{Zenit, R}, \au{Campbell, CS} \& \au{Brennen, CE}}
  \yr{2002}  \at{Revisiting the 1954 suspension experiments of ra bagnold}.
  \jt{J. Fluid Mech.}  \bvol{452},  \pg{1--24}.

\bibitem[Jiang {\em et~al.\/}(2020)Jiang, Calzavarini \&
  Sun]{jiang2020rotation}
{\sc \au{Jiang, Linfeng}, \au{Calzavarini, Enrico} \& \au{Sun, Chao}} \yr{2020}
   \at{Rotation of anisotropic particles in rayleigh--b{\'e}nard turbulence}.
  \jt{J. Fluid Mech.}  \bvol{901}.

\bibitem[Krieger \& Dougherty(1959)]{krieger1959mechanism}
{\sc \au{Krieger, Irvin~M} \& \au{Dougherty, Thomas~J}} \yr{1959}  \at{A
  mechanism for non-newtonian flow in suspensions of rigid spheres}.
  \jt{Trans. Soc. Rheol.}  \bvol{3}~(1),  \pg{137--152}.

\bibitem[Lashgari {\em et~al.\/}(2014)Lashgari, Picano, Breugem \&
  Brandt]{lashgari2014laminar}
{\sc \au{Lashgari, Iman}, \au{Picano, Francesco}, \au{Breugem, Wim-Paul} \&
  \au{Brandt, Luca}} \yr{2014}  \at{Laminar, turbulent, and inertial
  shear-thickening regimes in channel flow of neutrally buoyant particle
  suspensions}.  \jt{Phys. Rev. Lett.}  \bvol{113}~(25),  \pg{254502}.

\bibitem[Lathrop {\em et~al.\/}(1992)Lathrop, Fineberg \&
  Swinney]{lathrop1992transition}
{\sc \au{Lathrop, Daniel~P}, \au{Fineberg, Jay} \& \au{Swinney, Harry~L}}
  \yr{1992}  \at{Transition to shear-driven turbulence in couette-taylor flow}.
   \jt{Phys. Rev. A}  \bvol{46}~(10),  \pg{6390}.

\bibitem[Lohse(2018)]{lohse2018bubble}
{\sc \au{Lohse, Detlef}} \yr{2018}  \at{Bubble puzzles: From fundamentals to
  applications}.  \jt{Phys. Rev. Fluids}  \bvol{3}~(11),  \pg{110504}.

\bibitem[Lovecchio {\em et~al.\/}(2019)Lovecchio, Climent, Stocker \&
  Durham]{lovecchio2019chain}
{\sc \au{Lovecchio, Salvatore}, \au{Climent, Eric}, \au{Stocker, Roman} \&
  \au{Durham, William~M}} \yr{2019}  \at{Chain formation can enhance the
  vertical migration of phytoplankton through turbulence}.  \jt{Sci. Adv.}
  \bvol{5}~(10),  \pg{eaaw7879}.

\bibitem[Lu {\em et~al.\/}(2020)Lu, Xia, Shao \& Fu]{particledetection}
{\sc \au{Lu, Changsheng}, \au{Xia, Siyu}, \au{Shao, Ming} \& \au{Fu, Yun}}
  \yr{2020}  \at{Arc-support line segments revisited: An efficient high-quality
  ellipse detection}.  \jt{IEEE Trans. Image Process.}  \bvol{29},
  \pg{768--781}.

\bibitem[Magnaudet \& Eames(2000)]{magnaudet2000motion}
{\sc \au{Magnaudet, Jacques} \& \au{Eames, Ian}} \yr{2000}  \at{The motion of
  high-reynolds-number bubbles in inhomogeneous flows}.  \jt{Annu. Rev. Fluid
  Mech.}  \bvol{32}~(1),  \pg{659--708}.

\bibitem[Maryami {\em et~al.\/}(2014)Maryami, Farahat, Mayam {\em
  et~al.\/}]{maryami2014bubbly}
{\sc \au{Maryami, R}, \au{Farahat, S}, \au{Mayam, MH~Shafiei} \& \au{others}}
  \yr{2014}  \at{Bubbly drag reduction in a vertical couette--taylor system
  with superimposed axial flow}.  \jt{Fluid Dyn. Res.}  \bvol{46}~(5),
  \pg{055504}.

\bibitem[Mathai {\em et~al.\/}(2018)Mathai, Huisman, Sun, Lohse \&
  Bourgoin]{mathai2018dispersion}
{\sc \au{Mathai, Varghese}, \au{Huisman, Sander~G}, \au{Sun, Chao}, \au{Lohse,
  Detlef} \& \au{Bourgoin, Micka{\"e}l}} \yr{2018}  \at{Dispersion of air
  bubbles in isotropic turbulence}.  \jt{Phys. Rev. Lett.}  \bvol{121}~(5),
  \pg{054501}.

\bibitem[Mathai {\em et~al.\/}(2020)Mathai, Lohse \& Sun]{mathai2020bubbly}
{\sc \au{Mathai, Varghese}, \au{Lohse, Detlef} \& \au{Sun, Chao}} \yr{2020}
  \at{Bubbly and buoyant particle-laden turbulent flows}.  \jt{Annu. Rev.
  Condens. Matter Phys.}  \bvol{11},  \pg{529--559}.

\bibitem[Mittal {\em et~al.\/}(2020)Mittal, Ni \& Seo]{mittal2020flow}
{\sc \au{Mittal, Rajat}, \au{Ni, Rui} \& \au{Seo, Jung-Hee}} \yr{2020}  \at{The
  flow physics of covid-19}.  \jt{J. Fluid Mech.}  \bvol{894}.

\bibitem[Olivucci {\em et~al.\/}(2021)Olivucci, Wise \&
  Ricco]{olivucci2021reduction}
{\sc \au{Olivucci, Paolo}, \au{Wise, Daniel~J} \& \au{Ricco, Pierre}} \yr{2021}
   \at{Reduction of turbulent skin-friction drag by passively rotating discs}.
  \jt{J. Fluid Mech.}  \bvol{923}.

\bibitem[Park {\em et~al.\/}(2018)Park, O'Keefe \& Richter]{park2018rayleigh}
{\sc \au{Park, Hyungwon~John}, \au{O'Keefe, Kevin} \& \au{Richter, David~H}}
  \yr{2018}  \at{Rayleigh-b{\'e}nard turbulence modified by two-way coupled
  inertial, nonisothermal particles}.  \jt{Phys. Rev. Fluids}  \bvol{3}~(3),
  \pg{034307}.

\bibitem[Pedley \& Kessler(1992)]{pedley1992hydrodynamic}
{\sc \au{Pedley, TJ} \& \au{Kessler, John~O}} \yr{1992}  \at{Hydrodynamic
  phenomena in suspensions of swimming microorganisms}.  \jt{Annu. Rev. Fluid
  Mech.}  \bvol{24}~(1),  \pg{313--358}.

\bibitem[Peskin(2002)]{peskin2002immersed}
{\sc \au{Peskin, Charles~S}} \yr{2002}  \at{The immersed boundary method}.
  \jt{Acta Numer.}  \bvol{11},  \pg{479--517}.

\bibitem[Picano {\em et~al.\/}(2015)Picano, Breugem \&
  Brandt]{picano2015turbulent}
{\sc \au{Picano, Francesco}, \au{Breugem, Wim-Paul} \& \au{Brandt, Luca}}
  \yr{2015}  \at{Turbulent channel flow of dense suspensions of neutrally
  buoyant spheres}.  \jt{J. Fluid Mech.}  \bvol{764},  \pg{463--487}.

\bibitem[Picano {\em et~al.\/}(2013)Picano, Breugem, Mitra \&
  Brandt]{picano2013shear}
{\sc \au{Picano, Francesco}, \au{Breugem, Wim-Paul}, \au{Mitra, Dhrubaditya} \&
  \au{Brandt, Luca}} \yr{2013}  \at{Shear thickening in non-brownian
  suspensions: an excluded volume effect}.  \jt{Phys. Rev. Lett.}
  \bvol{111}~(9),  \pg{098302}.

\bibitem[Rosti \& Takagi(2021)]{rosti2021shear}
{\sc \au{Rosti, Marco~E} \& \au{Takagi, Shu}} \yr{2021}  \at{Shear-thinning and
  shear-thickening emulsions in shear flows}.  \jt{Phys. Fluids}
  \bvol{33}~(8),  \pg{083319}.

\bibitem[Rusconi {\em et~al.\/}(2014)Rusconi, Guasto \&
  Stocker]{rusconi2014bacterial}
{\sc \au{Rusconi, Roberto}, \au{Guasto, Jeffrey~S} \& \au{Stocker, Roman}}
  \yr{2014}  \at{Bacterial transport suppressed by fluid shear}.  \jt{Nat.
  Phys.}  \bvol{10}~(3),  \pg{212--217}.

\bibitem[Sanders {\em et~al.\/}(2006)Sanders, Winkel, Dowling, Perlin \&
  Ceccio]{sanders2006bubble}
{\sc \au{Sanders, Wendy~C}, \au{Winkel, Eric~S}, \au{Dowling, David~R},
  \au{Perlin, Marc} \& \au{Ceccio, Steven~L}} \yr{2006}  \at{Bubble friction
  drag reduction in a high-reynolds-number flat-plate turbulent boundary
  layer}.  \jt{J. Fluid Mech.}  \bvol{552},  \pg{353--380}.

\bibitem[Saw {\em et~al.\/}(2008)Saw, Shaw, Ayyalasomayajula, Chuang \&
  Gylfason]{saw2008inertial}
{\sc \au{Saw, Ewe~Wei}, \au{Shaw, Raymond~A}, \au{Ayyalasomayajula,
  Sathyanarayana}, \au{Chuang, Patrick~Y} \& \au{Gylfason, Armann}} \yr{2008}
  \at{Inertial clustering of particles in high-reynolds-number turbulence}.
  \jt{Phys. Rev. Lett.}  \bvol{100}~(21),  \pg{214501}.

\bibitem[Stickel \& Powell(2005)]{stickel2005fluid}
{\sc \au{Stickel, Jonathan~J} \& \au{Powell, Robert~L}} \yr{2005}  \at{Fluid
  mechanics and rheology of dense suspensions}.  \jt{Annu. Rev. Fluid Mech.}
  \bvol{37},  \pg{129--149}.

\bibitem[Tagawa {\em et~al.\/}(2013)Tagawa, Roghair, Prakash, van
  Sint~Annaland, Kuipers, Sun \& Lohse]{tagawa2013clustering}
{\sc \au{Tagawa, Yoshiyuki}, \au{Roghair, Ivo}, \au{Prakash, Vivek~N}, \au{van
  Sint~Annaland, Martin}, \au{Kuipers, Hans}, \au{Sun, Chao} \& \au{Lohse,
  Detlef}} \yr{2013}  \at{The clustering morphology of freely rising deformable
  bubbles}.  \jt{J. Fluid Mech.}  \bvol{721}.

\bibitem[Toschi \& Bodenschatz(2009)]{toschi2009lagrangian}
{\sc \au{Toschi, Federico} \& \au{Bodenschatz, Eberhard}} \yr{2009}
  \at{Lagrangian properties of particles in turbulence}.  \jt{Annu. Rev. Fluid
  Mech.}  \bvol{41},  \pg{375--404}.

\bibitem[Uhlmann(2008)]{uhlmann2008interface}
{\sc \au{Uhlmann, Markus}} \yr{2008}  \at{Interface-resolved direct numerical
  simulation of vertical particulate channel flow in the turbulent regime}.
  \jt{Phys. Fluids}  \bvol{20}~(5),  \pg{053305}.

\bibitem[Van~Gils {\em et~al.\/}(2011)Van~Gils, Huisman, Bruggert, Sun \&
  Lohse]{van2011torque}
{\sc \au{Van~Gils, Dennis~PM}, \au{Huisman, Sander~G}, \au{Bruggert, Gert-Wim},
  \au{Sun, Chao} \& \au{Lohse, Detlef}} \yr{2011}  \at{Torque scaling in
  turbulent taylor-couette flow with co-and counterrotating cylinders}.
  \jt{Phys. Rev. Lett.}  \bvol{106}~(2),  \pg{024502}.

\bibitem[Verschoof {\em et~al.\/}(2016)Verschoof, Van Der~Veen, Sun \&
  Lohse]{verschoof2016bubble}
{\sc \au{Verschoof, Ruben~A}, \au{Van Der~Veen, Roeland~CA}, \au{Sun, Chao} \&
  \au{Lohse, Detlef}} \yr{2016}  \at{Bubble drag reduction requires large
  bubbles}.  \jt{Phys. Rev. Lett.}  \bvol{117}~(10),  \pg{104502}.

\bibitem[Voth \& Soldati(2017)]{voth2017anisotropic}
{\sc \au{Voth, Greg~A} \& \au{Soldati, Alfredo}} \yr{2017}  \at{Anisotropic
  particles in turbulence}.  \jt{Annu. Rev. Fluid Mech.}  \bvol{49},
  \pg{249--276}.

\bibitem[Wang {\em et~al.\/}(2017{\natexlab{{\em a\/}}})Wang, Abbas \&
  Climent]{wang2017modulation}
{\sc \au{Wang, Guiquan}, \au{Abbas, Micheline} \& \au{Climent, {\'E}ric}}
  \yr{2017{\natexlab{{\em a\/}}}}  \at{Modulation of large-scale structures by
  neutrally buoyant and inertial finite-size particles in turbulent couette
  flow}.  \jt{Phys. Rev. Fluids}  \bvol{2}~(8),  \pg{084302}.

\bibitem[Wang {\em et~al.\/}(2017{\natexlab{{\em b\/}}})Wang, Sierakowski \&
  Prosperetti]{wang2017fully}
{\sc \au{Wang, Yayun}, \au{Sierakowski, Adam~J} \& \au{Prosperetti, Andrea}}
  \yr{2017{\natexlab{{\em b\/}}}}  \at{Fully-resolved simulation of particulate
  flows with particles--fluid heat transfer}.  \jt{J. Comput. Phys.}
  \bvol{350},  \pg{638--656}.

\bibitem[Wang {\em et~al.\/}(2019)Wang, Mathai \& Sun]{wang2019self}
{\sc \au{Wang, Ziqi}, \au{Mathai, Varghese} \& \au{Sun, Chao}} \yr{2019}
  \at{Self-sustained biphasic catalytic particle turbulence}.  \jt{Nat.
  Commun.}  \bvol{10}~(1),  \pg{1--7}.

\bibitem[Wang {\em et~al.\/}(2020)Wang, Mathai \& Sun]{wang2020experimental}
{\sc \au{Wang, Ziqi}, \au{Mathai, Varghese} \& \au{Sun, Chao}} \yr{2020}
  \at{Experimental study of the heat transfer properties of self-sustained
  biphasic thermally driven turbulence}.  \jt{Int. J. Heat Mass Transf.}
  \bvol{152},  \pg{119515}.

\bibitem[Will {\em et~al.\/}(2021)Will, Mathai, Huisman, Lohse, Sun \&
  Krug]{will2021kinematics}
{\sc \au{Will, Jelle~B}, \au{Mathai, Varghese}, \au{Huisman, Sander~G},
  \au{Lohse, Detlef}, \au{Sun, Chao} \& \au{Krug, Dominik}} \yr{2021}
  \at{Kinematics and dynamics of freely rising spheroids at high reynolds
  numbers}.  \jt{J. Fluid Mech.}  \bvol{912}.

\bibitem[Yi {\em et~al.\/}(2021)Yi, Toschi \& Sun]{yi2021global}
{\sc \au{Yi, Lei}, \au{Toschi, Federico} \& \au{Sun, Chao}} \yr{2021}
  \at{Global and local statistics in turbulent emulsions}.  \jt{J. Fluid Mech.}
   \bvol{912}.

\bibitem[Zade {\em et~al.\/}(2018)Zade, Costa, Fornari, Lundell \&
  Brandt]{zade2018experimental}
{\sc \au{Zade, S.}, \au{Costa, P.}, \au{Fornari, W.}, \au{Lundell, F.} \&
  \au{Brandt, L.}} \yr{2018}  \at{Experimental investigation of turbulent
  suspensions of spherical particles in a square duct}.  \jt{J. Fluid Mech.}
  \bvol{857},  \pg{748--783}.

\bibitem[Zhang \& Prosperetti(2010)]{zhang2010physics}
{\sc \au{Zhang, Quan} \& \au{Prosperetti, Andrea}} \yr{2010}  \at{Physics-based
  analysis of the hydrodynamic stress in a fluid-particle system}.  \jt{Phys.
  Fluids}  \bvol{22}~(3),  \pg{033306}.

\end{thebibliography}

\end{document}